\documentclass[11pt]{article}

\usepackage{amsmath}
\usepackage{amssymb}
\usepackage{amsthm}
\usepackage{tikz}
\usepackage{graphicx}
\usepackage{float}
\usepackage{algorithm}
\usepackage[noend]{algpseudocode}
\usepackage{tabularx}
\usepackage[margin=1in]{geometry}
\usepackage{arydshln}
\usepackage{multirow}
\usepackage{wrapfig}
\usepackage{url}
\usepackage{bold-extra}
\usepackage[margin=0.5cm]{caption}

\usetikzlibrary{positioning,calc,shapes,arrows}
\usetikzlibrary{backgrounds}
\usetikzlibrary{matrix,shadows,arrows} 
\usetikzlibrary{decorations.pathreplacing}

\def\etal.{et\penalty50\ al.}
\theoremstyle{plain}

\theoremstyle{definition}
\newtheorem{definition}{Definition}[section]

\theoremstyle{remark}
\newtheorem{question}{Question}[section]

\newtheorem{remark}[question]{Remark}
\newtheorem{openproblem}[question]{Open Problem}

\theoremstyle{plain}
\newtheorem*{theorem*}{Theorem}

\DeclareMathOperator{\OPT}{OPT}
\DeclareMathOperator{\ALG}{ALG}

\DeclareMathOperator{\Ima}{Im}
\DeclareMathOperator*{\argmin}{arg\,min}

\DeclareMathOperator{\Bin}{Bin}

\DeclareMathOperator{\rk}{rk}

\title{An Experimental Study of Algorithms for Online Bipartite Matching}
\author{Allan Borodin\thanks{Research is supported by NSERC.} \\ University of Toronto \\ \textsf{bor@cs.toronto.edu}  \and 
Christodoulos Karavasilis\footnotemark[1] \\ University of Toronto \\ \textsf{ckar@cs.toronto.edu} \and 
Denis Pankratov\footnotemark[1] \\ Concordia University \\ \textsf{denis.pankratov@concordia.ca}}
\date{\today}

\begin{document}

\maketitle

\begin{abstract}
We perform an experimental study of algorithms for online bipartite matching under the known i.i.d. input model with integral types. In the last decade, there has been substantial effort in designing complex algorithms with the goal of improving worst-case approximation ratios. Our goal is to determine how these algorithms perform on more practical instances rather than worst-case instances. In particular, we are interested in whether the ranking of the algorithms by their worst-case performance is consistent with the ranking of the algorithms by their average-case/practical performance. We are also interested in whether preprocessing times and implementation difficulties that are introduced by these algorithms are justified in practice. To that end we evaluate these algorithms on different random inputs as well as real-life instances obtained from publicly available repositories. We compare these algorithms against several simple greedy-style algorithms. Most of the complex algorithms in the literature are presented as being non-greedy (i.e., an algorithm can intentionally skip matching a node that has available neighbors) to simplify the analysis. Every such algorithm can be turned into a greedy one without hurting its worst-case performance. On our benchmarks, non-greedy versions of these algorithms perform much worse than their greedy versions. Greedy versions perform about as well as the simplest greedy algorithm by itself. This, together with our other findings, suggests that simplest greedy algorithms are competitive with the state-of-the-art worst-case algorithms for online bipartite matching on many average-case and practical input families. Greediness is by far the most important property of online algorithms for bipartite matching.
\end{abstract}

\section{Introduction}
\label{sec:intro}
One of the most active areas of theoretical computer science is the design and analysis of ``efficient'' approximation  algorithms.  
Often the objective is to establish the best approximation ratio achieved by a polynomial time algorithm. Such analysis is often done in terms of adversarial worst-case inputs,  or in the case of stochastic analysis, in terms of a worst-case i.i.d. distributional setting. However, such analysis can be and is challenged as to whether or not these worst-case approximation bounds reflect results for more ``realistic'' settings. There are many reasons for the perceived and observed gap between  theory and practice: asymptotic time bounds can hide large constant factors, typical inputs are not worst-case inputs, and simple algorithms are much easier to implement and are usually preferred (by practitioners)  over more complex algorithms.  

The most common approach to better understanding the gap between theory and practice is to perform experimental studies with respect to data that better reflects reality\footnote{This is not to say that the gap between theory and practice is restricted to experimental studies. Other approaches, such as smoothed analysis, as initiated in  \cite{SpielmanT09}, and perturbation stable instances, as initiated in  \cite{Bilul12}, have also been proposed. Thus far these insightful analytical approaches have not yet been widely accepted.  Arguably, experimental analysis remains the most common method for trying to understand the comparative performance  of algorithms.}. Following this approach we wish to study relatively simple greedy and ``greedy-like'' algorithms for online bipartite  matching in comparison with more complex non-greedy algorithms that have been designed for a known distribution stochastic setting. Since bipartite matching can be offline solved optimally and relatively efficiently, we are able to precisely compute the observed competitive ratios. We will consider the i.i.d. model (with integral types) where there is a known type graph with respect to which online nodes are drawn i.i.d. such that the expected number of occurrences is integral for each online node.  We consider both synthetically generated type graphs as well as some type graphs based on real-world applications. Our experimental study indicates that simple greedy and greedy-like algorithms (that are unaware of the type graph) perform quite well in terms of the observed competitive ratio when compared to the significantly more complex algorithms designed to exploit the given known type graph. That is, while the provable worst-case approximation ratios (in expectation over the distribution) of these non-greedy algorithms are much better than what can be achieved by the simple greedy like algorithms we consider, there is a good reason why practitioners might want to use simple greedy algorithms. The  more complicated algorithms for known type graphs are stated as  being non-greedy (in the sense that an online node is not necessarily matched whenever possible). However, we show that ``greediness'' can be easily achieved without loss of generality and, moreover,  greediness is necessary for any algorithm to achieve good performance in practice.

The remainder of the paper is organized as follows. In Section~\ref{sec:prelim} we describe the set of algorithms under consideration. This includes two simple greedy algorithms (namely, a a simple deterministic greedy algorithm and the randomized \textsc{Ranking} algorithm~\cite{KarpVV90}), and five state of the art algorithms for the known type graph model with integral types. Some of these algorithms have only been informally described  in the literature and we provide a more detailed description when needed. We also consider a linear time two-pass ``online'' algorithm~\cite{Durr2016}, which experimentally is almost a proxy for obtaining optimality.  In Section~\ref{sec:experimental_setup}, we discuss the data sets we use as well as the experimental setup. Section~\ref{sec:experimental_results} provides the experimental results in terms of the observed competitive ratio. We also provide some timing results verifying that indeed the simple linear time greedy algorithms are significantly faster than the algorithms designed for known type graphs. Finally, in Sections~\ref{sec:discussion} and \ref{sec:conclusion}, we summarize the experimental results drawing some overall conclusions from our experimental study.   

\section{Preliminaries}
\label{sec:prelim}
\label{sec:preliminaries}

We consider bipartite graphs $G=(L,R,E)$ with bi-partition $(L,R)$. We shall often refer to the nodes in $L$ as the left nodes, or the left-hand-side (LHS, for short) nodes, or the online nodes. Similarly, the nodes in $R$ are referred to as the right nodes, the right-hand-side (RHS) nodes, or the offline nodes.

In the online version of bipartite matching, the right side is known to the algorithm in advance. The left-hand-side nodes are revealed one-by-one in a given order. When an online node is revealed, all its neighbors are revealed as well. After each arrival of an online node, the algorithm makes an irrevocable decision on which neighbor to match the current online node (if at all).

\subsection{Definitions and Notation}

Let $M$ be a matching in a bipartite graph $G=(L,R,E)$. We say $\ell \in L$ \emph{participates} in the matching $M$ if there is $r \in R$ such that $\{ \ell, r\} \in M$. We write $M(\ell)$ to denote such $r$. If $\ell$ does not participate in $M$ then we define $M(\ell) := \bot$. The same notions are defined for $r \in R$ symmetrically.

We shall measure the performance of an algorithm in one of two ways: in terms of the observed asymptotic approximation ratio, or in terms of the fraction of the matched offline nodes.

\begin{definition}
Let $\ALG$ be an online algorithm (possibly randomized) solving the bipartite matching problem over random graphs $G_n$ parameterized by the input size $n = |R|$. We write $\ALG(G_n)$ to denote the expected size of the matching (random variable) that is constructed by running $\ALG$ on $G_n$. We write $\OPT(G_n)$ to denote the size of a maximum matching in $G_n$. \emph{The asymptotic approximation ratio} of $\ALG$ with respect to $G_n$ is defined as:
\[\rho(\ALG, G_n) = \liminf_{n \rightarrow \infty} \frac{\mathbb{E}(\ALG(G_n))}{\mathbb{E}(\OPT(G_n))}.\]
\emph{The fraction of matched offline nodes} of $\ALG$ with respect to $G_n$ is defined as:
\[\mu(\ALG, G_n) = \liminf_{n \rightarrow \infty} \frac{\mathbb{E}(\ALG(G_n))}{n}.\]
The expectations above are taken over the randomness of the algorithm and the randomness of the input.
\end{definition}

\subsection{Known I.I.D. Model and Integral Types}

In the known i.i.d. model, one first chooses a \emph{type graph} $G=(L,R,E)$ and a distribution $p: L \rightarrow [0,1]$ on the LHS nodes. In this case, the nodes in $L$ are also referred to as types. The type graph together with the distribution is given to the algorithm in advance. In the known i.i.d. model, an actual input instance $\hat{G}=(\hat{L}, R, \hat{E})$ is a random variable and is generated from $G$ as follows. The right hand side $R$ is the same in $G$ and $\hat{G}$, but the left-hand-side of $\hat{G}$ consists of $m$ i.i.d. samples from $p$. Thus, say a given node $\hat{\ell} \in \hat{L}$ has type $\ell \in L$, then the neighbors of $\hat{\ell}$ in $\hat{G}$ are the same as the neighbors of $\ell$ in $G$. The graph $\hat{G}$ is presented to the algorithm in the vertex arrival model (the order of vertices is the same as the order in which they were generated). Note that a particular type $\ell$ can be absent altogether or can be repeated a number of times in $\hat{G}$. We refer to $\hat{G}$ as the \emph{instance graph}. Note that the instance graph is fully specified by a pair $(G,v)$ where $G$ is a type graph and $v$ is a vector of types, i.e., $v \in L^m$. When $G$ is clear from the context, we will refer to $v$ as an instance. The probability of seeing a particular vector $v$ is given by $p(v) = \prod_{i = 1}^m p(v_i)$.

One common interpretation of the above model is in the world of online advertising. Let $R$ be the set of online advertisers, and $L$ be the set of user types. A user type essentially corresponds to a subset of advertisers that a user of that type might be interested in. Through past history, online advertising platforms have gathered statistical information about the rate of arrival of users of a particular type. Arrival of a user corresponds to an impression (i.e., a website banner) where any of the compatible advertisers can show their advertisement. Thus, in a simple version of the problem, an online platform utilizes past statistical information (the type graph and the distribution) to maximize the number of compatible advertisements shown (in the actual instance). In the literature, the nodes in $\hat{L}$ are often referred to as impressions and the nodes in $R$ are often referred to as advertisers. 

A known i.i.d. problem is said to have \emph{integral types} if the expected number of times a particular type occurs is integral. We will denote the number of times type $\ell$ occurs in an instance by the random variable $Z_\ell$. Then the condition of integral types is equivalent to $\mathbb{E}(Z_\ell) = p(\ell)m \in \mathbb{Z}$. While the parameters $|L|, |R|,$ and $m$ can all be different, the most common setting is $m = |L|$. This assumption together with integral types implies that without loss of generality one can take $p$ to be the uniform distribution on $L$ (by duplicating types as necessary). An additional common assumption is that $|L| = |R|$. In that case we talk about a single parameter $n = |L| = |R| = m$.

In our empirical evaluations, we only consider integral types, so when we say ``known i.i.d. model'' we mean the known i.i.d. model with integral types and uniform distribution, unless stated otherwise.

\subsection{Algorithms}

In this section we describe all algorithms that are included in our experimental study:

\begin{enumerate}
    \item \textsc{SimpleGreedy}.
    \item \textsc{Ranking} due to Karp et al.~\cite{KarpVV90}.
    \item \textsc{FeldmanEtAl} due to Feldman et al.~\cite{FeldmanMMM09}.
    \item \textsc{BahmaniKapralov} due to Bahmani and Kapralov~\cite{Bahmani2010}.
    \item \textsc{ManshadiEtAl} due to Manshadi et al.~\cite{Manshadi2011}.
    \item \textsc{JailletLu} due to Jaillet and Lu~\cite{Jaillet2014}.
    \item \textsc{BrubachEtAl} due to Brubach et al.~\cite{BrubachSSX16}.
    \item \textsc{Category-Advice} due to D\"urr et al.~\cite{Durr2016}.
    \item \textsc{3-Pass} due to Borodin et al~\cite{BorodinPS2018}.
    \item Offline optimal algorithm that runs Edmonds-Karp flow algorithm on the canonical flow network associated with a bipartite graph. Sometimes, we initialize the algorithm by a solution computed by one of the other algorithms.
\end{enumerate}
\begin{table}[]
    \centering
    \begin{tabular}{ |l||l|}
\hline
Algorithm & worst case analysis \\
\hline
\hline
\textsc{BrubachEtAl} & 0.7299 \cite{BrubachSSX16}\\
\hline
\textsc{JailletLu} & 0.7293 ($1-2/e^2$) \cite{Jaillet2014}\\
\hline
\textsc{ManshadiEtAl} & 0.7025 \cite{Manshadi2011}\\
\hline
\textsc{BahmaniKapralov} & 0.6990 \cite{Bahmani2010}\\
\hline
\textsc{Ranking} & 0.6961 \cite{Mahdian2011}  \\
\hline
\textsc{FeldmanEtAl} & 0.6702 ($\frac{1-2/e^2}{4/3 - 2/3e}$) \cite{FeldmanMMM09}\\
\hline
\textsc{Category-Advice} & 0.6321 ($1-1/e$) \cite{Durr2016}\\
\hline
\textsc{3-Pass} & 0.6321 ($1-1/e)$) \cite{BorodinPS2018}\\
\hline
\textsc{Greedy} & 0.6321 ($1-1/e$) \cite{KarpVV90}\\
\hline
\end{tabular}
    \caption{Algorithms with their respective provable competitive ratios.}
    \label{tab:algos_ranking}
\end{table}
We begin by presenting several algorithms that work in the online adversarial setting. This is followed by the description of algorithms that work in the known i.i.d. setting, and other algorithms that do not fit into online or known i.i.d. settings. Observe that algorithms that are designed for the online adversarial setting also work in the known i.i.d. setting --- they just ignore the side information, i.e., the type graph. 

\subsubsection{Algorithms for Online Adversarial Setting}
We start with a helper subroutine, which we call \textsc{GreedyWithPermutation}. This online algorithm accepts the RHS $R$ and a permutation $\pi$ of $R$. The rank of $r \in R$, denoted by $\rk_\pi(r)$, is the position of $r$ when in the arrangement of $R$ according to $\pi$. The \textsc{GreedyWithPermutation} algorithm matches each online node with an available neighbor of smallest rank (if there is at least one available neighbor). The pseudocode is presented in Algorithm~\ref{alg:greedy-with-permutation}.

\begin{algorithm}
\caption{A helper algorithm.}\label{alg:greedy-with-permutation}
\begin{algorithmic}
\Procedure{GreedyWithPermutation}{$G=(L,R,E), \pi : R \rightarrow R$}
\ForAll{$\ell \in L$}
\State When $\ell$ arrives, let $N(\ell)$ be the set of unmatched neighbors of $\ell$.
\If{$N(\ell) \neq \emptyset$}
\State Match $\ell$ with $\argmin \{ \rk_\pi(r) \mid r \in N(\ell) \}$.
\EndIf
\EndFor
\EndProcedure
\end{algorithmic}
\end{algorithm}

\noindent {\bf \large \textsc{SimpleGreedy}.} Next, we describe the simplest online algorithm -- \textsc{SimpleGreedy}. The \textsc{SimpleGreedy} algorithm is obtained by fixing a permutation $\pi$ on the RHS and applying \textsc{GreedyWithPermutation}. While $\pi$ could be any fixed permutation (not depending on the type graph), for concreteness, we define it to be the following. The RHS nodes are labelled with strings over some alphabet. We define $\pi_\text{alphabet}$ to be the ordering of the RHS nodes alphabetically according to their labels. Thus, formally \textsc{SimpleGreedy}$(G)$=\textsc{GreedyWithPermutation}$(G,\pi_\text{alphabet})$.

\begin{remark}
\label{rem:greedy}
A word of caution with regards to the terminology: \textsc{SimpleGreedy} should not be confused with an arbitrary \emph{greedy} algorithm. When we say that an algorithm is greedy, we mean that it has the following property: whenever a given online node has at least one unmatched neighbor, this online node is guaranteed to be matched. This property alone is not sufficient to specify the algorithm, since the algorithm also needs to break ties when several unmatched neighbors are available. \textsc{SimpleGreedy} is a very specific greedy algorithm, which breaks ties according to an alphabetical order. It turns out that \emph{any algorithm} for online bipartite matching can be turned into a greedy one without hurting its approximation ratio. In particular, without loss of generality, an optimal algorithm is greedy. Thus, the whole area of designing good online algorithms for bipartite matching revolves around designing better and better tie-breaking rules. We discuss this in more details below when we talk about more advanced algorithms for the known i.i.d model.
\end{remark}

\noindent {\bf \large \textsc{Ranking}.} The next algorithm is \textsc{Ranking} due to Karp et al.~\cite{KarpVV90}. Unlike the previous algorithms, \textsc{Ranking} is randomized. Let $S_R$ denote the set of all permutation of the RHS $R$. \textsc{Ranking} samples $\pi$ uniformly at random from $S_R$ prior to seeing any online nodes. This is followed by running \textsc{GreedyWithPermutation} with $\pi$ as the input permutation --- see Algorithm~\ref{alg:ranking}.

\begin{algorithm}
\caption{A randomized algorithm due to Karp et al.~\cite{KarpVV90}.}\label{alg:ranking}
\begin{algorithmic}
\Procedure{Ranking}{$G=(L,R,E)$}
\State Sample a permutation $\pi: R \rightarrow R$ uniformly at random.
\State Run \textsc{GreedyWithPermutation}$(G,\pi)$.
\EndProcedure
\end{algorithmic}
\end{algorithm}

\subsubsection{Algorithms for Known I.I.D. Setting}

We start with a special subroutine. Consider a bipartite graph of maximum degree $2$, i.e., a set of paths and cycles. Such a graph can be decomposed into two matchings, which we will call blue and red. Strictly speaking, the blue subgraph returned by the subroutine is not always a matching;  sometimes it is a matching plus some extra edges. However, the blue subgraph always satisfies the property that there is at most one edge incident on each LHS node, i.e., the blue subgraph is a ``matching on the left.'' For simplicity and slightly abusing notation, we shall sometimes refer to both blue and red subgraphs as matchings. However, for clarity, we can say that the blue edges form a ``semi-matching''. When we actually run the Feldman et al algorithm on an i.i.d. instance, the blue edges become a matching as determined by the assignment of the online node.  We present a particular decomposition in Algorithm~\ref{alg:blue-red-decomposition}, which we call \textsc{BlueRedDecomposition} and which is due to Feldman et al.~\cite{FeldmanMMM09}. This decomposition is used in several algorithms that we consider later.

\begin{algorithm}
\caption{Blue red decomposition due to Feldman et al.~\cite{FeldmanMMM09}. Applies to bipartite graphs of maximum degree 2.}\label{alg:blue-red-decomposition}
\begin{algorithmic}
\Procedure{BlueRedDecomposition}{$G=(L,R,E)$}
\State Color edges of the cycles alternating blue and red.
\State Color edges of the odd-length paths alternating blue and red, with more blue than red.
\State For the even-length paths that start and end with nodes in $R$, alternate blue and red.
\State For the even length paths that start and end with nodes in $L$, color the first two edges blue, then alternate red, blue, red, blue, etc.
\State \Return (semi-matching formed by blue edges, matching formed by red edges).
\EndProcedure
\end{algorithmic}
\end{algorithm}

\noindent {\bf \large \textsc{FeldmanEtAl}.} The first algorithm to ever beat the $1-1/e$ barrier of the online adversarial model in the known i.i.d. model is due to Feldman et al.~\cite{FeldmanMMM09}. The algorithm has a preprocessing stage and the online stage. In the preprocessing stage, the algorithm solves the following modification of the standard network flow problem for biparite matching: add two new nodes $s$ and $t$, add directed edges each from $s$ to $r$ for each $r \in R$, and add directed edges from $\ell$ to $t$ for each $\ell \in L$, orient the rest of the edges in $G$ from RHS to LHS (these edges will be called the graph edges). Each outgoing edge from $s$, as well as each incoming edge into $t$, has capacity 2. The rest of the edges have capacities 1. We denote this flow network by $\widetilde{G}$. The algorithm of Feldman et al. finds an integral optimal solution to this network flow problem. The subgraph induced by the graph edges with positive flow on them has maximum degree 2. The last step of the preprocessing stage is to apply \textsc{BlueRedDecomposition} to this subgraph to obtain a blue semi-matching $M_b$ and a red matching $M_r$. In the online stage, the algorithm receives online nodes in the i.i.d. fashion and matches them as follows: if a node of type $i$ arrives for the first time, the algorithm tries to match it to $M_b(i)$. If $M_b(i) = \bot$ or $M_b(i)$ has been previously matched, the algorithm leaves the current node unmatched. If a node of type $i$ arrives for the second time, the algorithm tries to match it to $M_r(i)$. Otherwise, a node of type $i$ is left unmatched. See Algorithm~\ref{alg:feldman} for the pseudocode.

\begin{algorithm}
\caption{The known i.i.d. algorithm of Feldman et al.~\cite{FeldmanMMM09}.}\label{alg:feldman}
\begin{algorithmic}
\Procedure{FeldmanEtAl}{$G=(L,R,E)$ -- type graph}
\State \Comment{Preprocessing stage:}
\State Set up flow network $\widetilde{G}=(\widetilde{V}, \widetilde{E})$, where 
\State $\widetilde{V} = L \cup R \cup \{s, t\}$
\State $\widetilde{E} = \{(s,r) \mid r \in R\} \cup \{(\ell, t) \mid \ell \in L\} \cup \{(r, \ell) \mid \{r, \ell\} \in E\}$.
\State Set up capacities $cap(s,r) = 2, cap(\ell, t) = 2$ for $\ell \in L, r \in R$ and $cap(r, \ell) = 1$ for $(r, \ell) \in \widetilde{E}$.
\State Solve the flow network to obtain a maximum integral flow $f$.
\State Let $G'$ denote the bipartite subgraph induced by edges $\{ r, \ell\}$ such that $f(r, \ell) = 1$.
\State Set $(M_b, M_r) = \textsc{BlueRedDecomposition}(G')$.
\State \Comment{Online stage:}
\ForAll{arriving online nodes $u$}
\State Let $\ell$ denote the type of $u$.
\If{it is the first arrival of type $\ell$ and $M_b(\ell) \neq \bot$ and $M_b(\ell)$ is unmatched}
\State Match $u$ to $M_b(\ell)$.
\EndIf
\If{it is the second arrival of type $\ell$ and $M_b(\ell) \neq \bot$ and $M_b(\ell)$ is unmatched}
\State Match $u$ to $M_r(\ell)$.
\EndIf
\EndFor
\EndProcedure
\end{algorithmic}
\end{algorithm}

\noindent {\bf \large \textsc{BahmaniKapralov}.} Bahmani and Kapralov~\cite{Bahmani2010} observed that the performance of Feldman et al. algorithm can be improved by modifying the preprocessing stage. Recall, that $G$ refers to the type graph, $\widetilde{G}$ to the associated flow network in \textsc{FeldmanEtAl}, and $f$ is an integral max flow in $\widetilde{G}$. Consider a subset $A$ of $L$ and define $A^z$ to be those vertices in $A$ such that the amount of flow through them in $f$ is $z$ for $z \in \{0,1,2\}$. In other words, no flow goes through vertices in $A^0$, one unit of flow goes through each vertex in $A^1$, and two units of flow go through each vertex in $A^2$. The main insight of Bahmani and Kapralov is that the more balanced the flow is the better, i.e., we want $A^1$ to be as large as possible. They give a procedure that redirects some of the flow from $A^2$ into $A^0$ without affecting the optimality of the flow. The procedure actually works on two sets $A \subseteq L$ and $B \subseteq R$ and can be done to balance the flow either on the left or on the right. We first describe the procedure and then show which sets to apply it to in order to improve on the algorithm of Feldman et al. 

Lets first define the procedure to balance the left side (the right side can be handled similarly). The algorithm sets up a completely new flow network $\widehat{G}$ as follows: the vertex set of the network consists of $A \cup B$ together with two new vertices $s_A$ and $t_A$. We add an edge $(s_A, a)$ of unit capacity for each $a \in A^2$ and an edge $(a, t_A)$ of unit capacity for each $a \in A^0$. For each edge $(b, a)$ such that $a \in A, b \in B$ and $f(b,a) = 1$, we add an edge $(a,b)$ to the flow network of unit capacity (note that this essentially reverses the edges with positive flow in $\widetilde{G}$). For each $(b,a)$ in $\widetilde{G}$ with $f(b,a) = 0$ (in $\widetilde{G}$), we add an edge $(b,a)$ to $\widehat{G}$ of unit capacity (note that this essentially preserves the graph edges in $\widetilde{G}$ that do not carry any flow). Let $f_A$ denote an integral maximum flow in the newly constructed flow network. If we use the convention that $f(a,b) = - f(b,a)$ then by adding $f_A$ to $f$ on edges $(b,a)$ and fixing the flow on edges $(a,t)$ accordingly, we essentially ``undo'' some flow going into $A^2$ nodes and replace it with a flow going into $A^0$ nodes in $\widetilde{G}$. 

Perhaps, this is best illustrated with a small example. Consider $K_{4,2}$ type graph, where $L = \{\ell_1, \ell_2, \ell_3, \ell_4\}$ and $R = \{r_1, r_2\}$. One possible max flow that Feldman et al. algorithm finds for the corresponding network is to send two units of flow through $r_1$ and into $\ell_1, \ell_2$ and to send two units of flow through $r_2$ and into $\ell_1, \ell_2$. Call this flow $f$. Consider $A = L$ and $B = R$. Then $A^2 = \{\ell_1, \ell_2\}$ and $A^0 = \{\ell_3, \ell_4\}$. Solving the new flow problem corresponding to the balancing procedure we find that we can send one unit of flow through $\ell_1$ to $r_1$ and to $\ell_3$ and another unit of flow through $\ell_2$ to $r_2$ and to $\ell_4$. Thus, this new flow can be used to augment $f$: it undoes one unit of flow from $r_1$ to $\ell_1$ and replaces it with one unit of flow from $r_1$ to $\ell_3$, and it undoes one unit of flow from $r_2$ to $\ell_2$ and replaces it with one unit of flow from $r_2$ to $\ell_4$. This results in a new flow being completely balanced on the LHS, i.e., $A^1 = L$. The two balancing procedures are described in Algorithms~\ref{alg:balance-left} and~\ref{alg:balance-right}.

\begin{algorithm}
\caption{The balancing procedure on the LHS due to Bahmani and Kapralov~\cite{Bahmani2010} that takes as input type graph $G$, a maximum integral flow for the flow network from Feldman et al. and two sets $A \subseteq L$ and $B \subseteq R$. Returns a flow $f_A$ in the new flow network that can be used to define a more balanced $f$.}\label{alg:balance-left}
\begin{algorithmic}
\Procedure{BalanceLeft}{$G=(L,R,E), A, B, f$}
\State Set up flow network $\widehat{G}_A=(\widehat{V}_A, \widehat{E}_A)$, where 
\State $\widehat{V}_A = A \cup B \cup \{s_A, t_A\}$
\State $\widehat{E}_A = \{(s_A,a) \mid a \in A^2\} \cup \{(a, t_A) \mid a \in A^0\} \cup \{(a, b) \mid f(b,a) = 1\} \cup \{(b,a) \mid f(b,a) = 0\}$.
\State Set capacities of all edges in $\widehat{E}_A$ to 1.
\State Solve the flow network to obtain a maximum integral flow $f_A$.
\State \Return $f_A$.
\EndProcedure
\end{algorithmic}
\end{algorithm}

\begin{algorithm}
\caption{The balancing procedure on the RHS due to Bahmani and Kapralov~\cite{Bahmani2010} that takes as input type graph $G$, a maximum integral flow for the flow network from Feldman et al. and two sets $A \subseteq L$ and $B \subseteq R$. Returns a flow $f_A$ in the new flow network that can be used to define a more balanced $f$.}\label{alg:balance-right}
\begin{algorithmic}
\Procedure{BalanceRight}{$G=(L,R,E), A, B, f$}
\State Set up flow network $\widehat{G}_B=(\widehat{V}_B, \widehat{E}_B)$, where 
\State $\widehat{V}_B = A \cup B \cup \{s_B, t_B\}$
\State $\widehat{E}_B = \{(s_B,b) \mid b \in B^0\} \cup \{(b, t_B) \mid b \in B^2\} \cup \{(a, b) \mid f(b,a) = 1\} \cup \{(b,a) \mid f(b,a) = 0\}$.
\State Set capacities of all edges in $\widehat{E}_B$ to 1.
\State Solve the flow network to obtain a maximum integral flow $f_B$.
\State \Return $f_B$.
\EndProcedure
\end{algorithmic}
\end{algorithm}

Now, let $(S,T)$ be the min cut in the flow network $\widetilde{G}$ obtained in a standard way: $S$ is defined to be the set of nodes reachable from $s$ in the residual network defined by max flow $f$. Define $S_L = S \cap L, S_R = S \cap R, T_L = T \cap L, T_R = T \cap R$. Bahmani and Kapralov algorithm computes $f_L = \textsc{BalanceLeft}(G, T_L, T_R, f)$ and $f_R = \textsc{BalanceRight}(G,S_L,S_R,f)$. It then creates a subgraph of $G$ consisting of those edges $\{r, \ell\}$ that have $f(r, \ell) + f_L(r, \ell) + f_R(r, \ell) > 0$ (using the convention $f(u,v) = -f(v,u)$). The rest is exactly as in Feldman et al. --- use \textsc{BlueRedDecomposition} on this subgraph and use the resulting blue and red matchings in the online stage in the same way as Feldman et al. See Algorithm~\ref{alg:bahmani-kapralov} for the pseudocode.

\begin{algorithm}
\caption{The known i.i.d. algorithm of Bahmani and Kapralov~\cite{Bahmani2010}.}\label{alg:bahmani-kapralov}
\begin{algorithmic}
\Procedure{BahmaniKapralov}{$G=(L,R,E)$ -- type graph}
\State \Comment{Preprocessing stage:}
\State Compute $f$ as in Algorithm~\ref{alg:feldman}.
\State Compute the canonical $(S,T)$ cut from $f$.
\State Set $f_L = \textsc{BalanceLeft}(G, T_L, T_R, f)$.
\State Set $f_R = \textsc{BalanceRight}(G,S_L,S_R,f)$.
\State Let $G'$ be induced by edges $\{ r, \ell\}$ such that $f(r, \ell) + f_L(r, \ell) + f_R(r, \ell) > 0$ (using the convention $f(u,v) = -f(v,u)$). 
\State Set $(M_b, M_r) = \textsc{BlueRedDecomposition}(G')$.
\State \Comment{Online stage:}
\State Same as in Algorithm~\ref{alg:feldman}.
\EndProcedure
\end{algorithmic}
\end{algorithm}

\noindent {\bf \large \textsc{ManshadiEtAl}.} The next algorithm is due to Manshadi et al.~\cite{Manshadi2011} and it is based on the idea of a fractional optimal solution. Fix an algorithm for obtaining an offline optimal solution (e.g. Edmonds-Karp). Consider all possible instances arising out of the given type graph  $G = (L,R,E)$. Recall, that an instance can be described as a vector of types $v \in L^n$. We assume without loss of generality that the expected number of arrivals of nodes of a given type $\ell$ is bounded above by 1. The matching $M$ given by the optimal algorithm can be viewed as an indicator vector of length $|E|$ indexed by edge names. This indicator vector specifies for each position $\{\ell, r\}$ whether $\{\ell, r\}$ is in $M$ or not. Abusing the notation, we denote this indicator vector by $\OPT(v)$. An optimal fractional solution is given by the expected value of this indicator vector, i.e., $f_{\OPT} = \sum_{v \in L^n} p(v) \OPT(v)$. Observe that $f_{\OPT} \in [0,1]^E$ and for each edge $\{\ell, r\} \in E$ we have $f_{\OPT}(\{\ell, r\}) = $ the probability that edge $\{\ell, r\}$ appears in an optimal matching. 

Let $W_{\{\ell, r\}}$ denote the random variable indicating the event that $\{\ell, r\}$ appears in an optimal matching. Let $Z_\ell$ denote the number of online nodes generated of type $\ell$. For each $\ell$ we have $\sum_{r : \{\ell, r\} \in E} W_{\{\ell, r\}} \le Z_\ell$. By taking the expectation of both sides, we have $\sum_{r : \{\ell, r\} \in E} f_{\OPT}(\{\ell, r\}) \le 1$ (using the assumption described above). For a given type $\ell$ let $r_1, \ldots, r_k$ be its neighbors in $G$ ordered such that $f_{\OPT}(\{\ell, r_1\}) \ge f_{\OPT}(\{\ell, r_2\}) \ge \cdots \ge f_{\OPT}(\{\ell, r_k\})$. Add a dummy node $r_{k+1}$ and define $f_{\OPT}(\{\ell, r_{k+1}\}) = 1 - \sum_{i=1}^k f_{\OPT}(\{\ell, r_i\})$. The dummy node simulates the event that $\ell$ is not matched in an optimal solution --- for the purpose of the algorithm, the dummy node is always considered to be matched before the online stage. 

Now, $f_{\OPT}(\{\ell, \cdot\})$ defines a probability mass function (PMF) on the neighbors of $\ell$. The algorithm of Manshadi et al. samples two random neighbors from this distribution during the online stage in the following correlated fashion. Partition the interval $[0,1]$ into $k+1$ consecutive non overlapping intervals $I_p$ where the length of $I_p$ is $f_{\OPT}(\{\ell, r_p\})$. We denote this partition by $\mathcal{I}_\ell$. Also, partition the interval $[0,1]$ into $k+1$ consecutive non overlapping intervals $J_p$ where the length of $J_p$ is $f_{\OPT}(\{\ell, r_{p+1}\})$ if $p \le k$ and the length of $J_{k+1}$ is $f_{\OPT}(\{\ell, r_1\})$. We denote this partition by $\mathcal{J}_\ell$. In order to sample from the PMF on the neighbors of $\ell$, one could sample a uniform random number between $0$ and $1$ and output the neighbor of $\ell$ corresponding to the interval to which the number belongs. If we do this procedure independently for $I$ intervals and $J$ intervals, we get two independent samples. Instead, Manshadi et al. do the correlated sampling --- a single number is sampled between $0$ and $1$. Let $I_p$ and $J_q$ be the intervals in which this number falls. The two neighbors returned by the procedure are the two neighbors of $\ell$ corresponding to $I_r$ and $J_q$. The partitioning of $[0,1]$ into $I$ intervals and $J$ intervals was chosen so that there is as little overlap between intervals corresponding to the same neighbor as possible.

When an online node of type $\ell$ arrives, the algorithm of Manshadi et al. performs a correlated sampling from $\mathcal{I}_\ell$ and $\mathcal{J}_\ell$ as described above. Let $r_{\ell, 1}$ and $r_{\ell, 2}$ denote the two samples returned by the correlated sampling procedure. The algorithm tries to match the online node first to $r_{\ell, 1}$. If $r_{\ell, 1}$ was matched previously, the algorithm tries to match the online node to $r_{\ell, 2}$. If $r_{\ell, 2}$ was matched previously, the algorithm gives up on matching the online node. See Algorithm~\ref{alg:manshadi} for the pseudocode. There is an outstanding issue of how to compute $f_{\OPT}$ in practice. This is a difficult problem, and rather than computing it exactly, Manshadi et al. suggest approximating it by the Monte Carlo method -- sample a number of instances, solve them optimally, record the fraction of times each edge appears in an optimal offline solution. This is what we do in our implementation, as well (see Section~\ref{sec:implmentation-notes}).

\begin{algorithm}
\caption{The known i.i.d. algorithm of Manshadi et al.~\cite{Manshadi2011}.}\label{alg:manshadi}
\begin{algorithmic}
\Procedure{ManshadiEtAl}{$G=(L,R,E)$ -- type graph}
\State \Comment{Preprocessing stage:}
\State Compute a fractional optimal matching $f_{\OPT}$.
\State For each $\ell$ construct the two partitions $\mathcal{I}_\ell$ and $\mathcal{J}_\ell$.
\State \Comment{Online stage:}
\ForAll{arriving online nodes $u$}
\State Let $\ell$ denote the type of $u$.
\State Let $r_{\ell, 1}$ and $r_{\ell, 2}$ be the two neighbors of $\ell$ returned by the correlated sampling procedure performed on $\mathcal{I}_\ell$ and $\mathcal{J}_\ell$ as described in the text.
\If{$r_{\ell, 1}$ is unmatched}
\State Match $u$ to $r_{\ell, 1}$.
\ElsIf{$r_{\ell,2}$ is unmatched}
\State Match $u$ to $r_{\ell,2}$.
\Else
\State Leave $u$ unmatched.
\EndIf
\EndFor
\EndProcedure
\end{algorithmic}
\end{algorithm}

\noindent {\bf \large \textsc{JailletLu}.} Jaillet and Lu~\cite{Jaillet2014} introduced a template of algorithms called Random Lists Algorithms, RLA for short, for online bipartite matching under known i.i.d. input model. For type $\ell$, define $\Omega_\ell$ to be the set of all possible ordered (sub)lists of neighbors of $\ell$ in the type graph. In the preprocessing stage, an RLA constructs a distribution $D_\ell$ on $\Omega_\ell$ for each $\ell \in L$. In the online stage, when a node of type $\ell$ arrives, the RLA samples a list of neighbors from $D_\ell$ and matches the online node to the first available neighbor according to that list. If there are no available neighbors in that list, the online node is left unmatched. The pseudocode for this template appears in Algorithm~\ref{alg:rla}. In order to get an actual algorithm out of this template, one has to specify how $D_\ell$ are constructed in the preprocessing step.

\begin{algorithm}
\caption{Random Lists Algorithm template due to Jaillet and Lu~\cite{Jaillet2014}.}\label{alg:rla}
\begin{algorithmic}
\Procedure{RLA}{$G=(L,R,E)$ -- type graph}
\State \Comment{Preprocessing stage:}
\State For each $\ell \in L$ construct a distribution $D_\ell$ on $\Omega_\ell$.
\State \Comment{Online stage:}
\ForAll{arriving online nodes $u$}
\State Let $\ell$ denote the type of $u$.
\State Sample a list of neighbors of $\ell$ from $\Omega_\ell$ according to $D_\ell$.
\If{all neighbors in the list are matched}
\State Leave $u$ unmatched.
\Else
\State Match $u$ to the first available neighbor in the list.
\EndIf
\EndFor
\EndProcedure
\end{algorithmic}
\end{algorithm}

Jaillet and Lu~\cite{Jaillet2014} also gave an actual algorithm based on this template, which we refer to as \textsc{JailletLu}. Jaillet and Lu consider the following LP:

\begin{equation}
\label{eq:jailletlu}
\begin{array}{rll}
\text{maximize} & \sum_{\ell \in L, r \in R} f_{\ell, r} & \\
\text{subject to} & \sum_{\ell : \{\ell, r\} \in E} f_{\ell, r} \le 1 & r \in R \\
 & \sum_{r : \{\ell, r\} \in E} f_{\ell, r} \le 1 & \ell \in L \\
 & f_{\ell, r} \in [0, 2/3] & \ell \in L, r \in R, \{\ell, r\} \in E
\end{array}
\end{equation}

A vertex solution $f^*$ to this LP has the property that $f^*_{\ell, r} \in \{0, 1/3, 2/3\}$ for all $\ell \in L, r \in R$. Restrict the neighbors of $\ell$ to only those $r$ that have $f^*_{\ell, r} > 0$. There can be at most 3 neighbors, since for such $r$ we have $f^*_{\ell, r} \ge 1/3$. If $\sum_{r: \{\ell, r\} \in E} f^*_{\ell, r} < 1$, then add a dummy node $d_\ell$ and define $f^*_{\ell, d_\ell} = 1 - \sum_{r: \{\ell, r\} \in E} f^*_{\ell, r}$. Even after adding dummy nodes, each $\ell$ has at most 3 neighbors. Jaillet and Lu define $D_\ell$ such that it is supported only on lists of these restricted neighborhoods. More specifically, if $\ell$ has a single neighbor then $D_\ell$ assigns unit weight to the list consisting of that neighbor; if $\ell$ has two neighbors $r_1, r_2$ then $D_\ell$ assigns probability $f^*_{\ell, r_1}$ to the list $\langle r_1, r_2 \rangle$ and probability $f^*_{\ell, r_2}$ to the list $\langle r_2, r_1 \rangle$; if $\ell$ has three neighbors $r_1, r_2, r_3$ then $D_\ell$ assigns probability $1/6$ to each permutation of $r_1, r_2, r_3$. After that, \textsc{JailletLu} runs RLA with these distributions.

\noindent {\bf \large \textsc{BrubachEtAl}.} Next we describe the state-of-the-art\footnote{The state-of-the-art is in terms of the best provable competitive ratio over worst-case type graphs.} algorithm for the known i.i.d. input model with integral arrival rates due to Brubach et al.~\cite{BrubachSSX16}. This algorithm is (predictably) the most difficult to explain and implement. It is a RLA-syle algorithm. The preprocessing stage consists of five steps:

\begin{enumerate}
    \item solve a special LP,
    \item round the solution,
    \item apply the first modification to the rounded solution,
    \item apply the second modification to the modified solution from the second step,
    \item define distributions $D_\ell$ on $\Omega_\ell$ for each $\ell \in L$.
\end{enumerate}

Next, we describe each of these steps in detail. \textbf{Step 1} --- Brubach et al. define and solve the following LP:

\begin{equation}
\label{eq:brubachetal}
\begin{array}{rll}
\text{maximize} & \sum_{\ell \in L, r \in R} f_{\ell, r} & \\
\text{subject to} & \sum_{\ell : \{\ell, r\} \in E} f_{\ell, r} \le 1 & r \in R \\
 & \sum_{r : \{\ell, r\} \in E} f_{\ell, r} \le 1 & \ell \in L \\
 & 0 \le f_{\ell, r} \le 1-\frac{1}{e} & \ell \in L, r \in R, \{\ell, r\} \in E \\
 & f_{\ell_1, r} + f_{\ell_2, r} \le 1 - \frac{1}{e^2} & \ell_1, \ell_2 \in L, r \in R, \{\ell_1, r\}, \{\ell_2, r\} \in E
\end{array}
\end{equation}

The idea behind LP~\eqref{eq:brubachetal} is to introduce extra constraints to bring the optimal value of the objective down closer to the fractional optimal solution, while maintaining feasibility of the fractional optimal solution. Let $f^*$ denote an optimal solution to \eqref{eq:brubachetal}. \textbf{Step 2} is to apply the rounding procedure of Gandhi et al.~\cite{GandhiKPS2006} to $3f^*$, i.e., $f^*$ multiplicatively scaled by 3. This results in an integral vector $\widetilde{f}$ such that $\widetilde{f}_{\ell,r} \in \{0,1,2,3\}$. Then, Brubach et al. scale the rounded solution back down and set $h := \widetilde{f}/3$. For completeness, we describe the rounding procedure here. Say an edge in our bipartite graph is fractional if $f^*_{\ell, r} \not\in \mathbb{Z}$. While there are fractional edges remaining, repeat the following. Find either a cycle or a maximal path consisting only of fractional edges. Let $P$ denote this cycle/path, and partition it into two matchings $M_1$ and $M_2$. Define
\[ \alpha = \min \left\{\gamma > 0 \mid (\exists (i,j) \in M_1 : f^*_{i,j} + \gamma = \lceil f^*_{i,j} \rceil) \wedge (\exists (i,j) \in M_2 : f^*_{i,j}-\gamma = \lfloor f^*_{i,j} \rfloor)\right\}\]
\[ \beta = \min \left\{\gamma > 0 \mid (\exists (i,j) \in M_1 : f^*_{i,j} - \gamma = \lfloor f^*_{i,j} \rfloor) \wedge (\exists (i,j) \in M_2 : f^*_{i,j}+\gamma = \lceil f^*_{i,j} \rceil)\right\}.\]
With probability $\beta/(\alpha+\beta$ round $f^*_{i,j}$ to $f^*_{i,j} + \alpha$ for all $\{i,j\} \in M_1$ and to $f^*_{i,j}-\alpha$ for all $\{i, j\} \in M_2$. With complementary probability, round $f^*_{i,j}$ to  $f^*_{i,j} -\beta$ for all $\{i,j\} \in M_1$ and to $f^*_{i,j}+\beta$ for all $\{i, j\} \in M_2$

\begin{figure}[ht]
\centering

\begin{tikzpicture}[scale=0.75]

\node at (-4,1) {$(C_1)$};
\node at (2,1) {$(C_2)$};
\node at (8,1) {$(C_3)$};

\node[draw=black,circle,fill=white] (c1l1) at (-6,0) {$\ell_1$};
\node[draw=black,circle,fill=white] (c1l2) at (-6,-2) {$\ell_2$};

\node[draw=black,circle,fill=white] (c1r1) at (-2,0) {$r_1$};
\node[draw=black,circle,fill=white] (c1r2) at (-2,-2) {$r_2$};

\node[draw=black,circle,fill=white] (c2l1) at (0,0) {$\ell_1$};
\node[draw=black,circle,fill=white] (c2l2) at (0,-2) {$\ell_2$};

\node[draw=black,circle,fill=white] (c2r1) at (4,0) {$r_1$};
\node[draw=black,circle,fill=white] (c2r2) at (4,-2) {$r_2$};

\node[draw=black,circle,fill=white] (c3l1) at (6,0) {$\ell_1$};
\node[draw=black,circle,fill=white] (c3l2) at (6,-2) {$\ell_2$};

\node[draw=black,circle,fill=white] (c3r1) at (10,0) {$r_1$};
\node[draw=black,circle,fill=white] (c3r2) at (10,-2) {$r_2$};

\draw[thick,->] (2, -2.5) -- (2, -4.5);

\node[draw=black,circle,fill=white] (c2pl1) at (0,-5) {$\ell_1$};
\node[draw=black,circle,fill=white] (c2pl2) at (0,-7) {$\ell_2$};

\node[draw=black,circle,fill=white] (c2pr1) at (4,-5) {$r_1$};
\node[draw=black,circle,fill=white] (c2pr2) at (4,-7) {$r_2$};

\draw[thick,->] (8, -2.5) -- (8, -4.5);

\node[draw=black,circle,fill=white] (c3pl1) at (6,-5) {$\ell_1$};
\node[draw=black,circle,fill=white] (c3pl2) at (6,-7) {$\ell_2$};

\node[draw=black,circle,fill=white] (c3pr1) at (10,-5) {$r_1$};
\node[draw=black,circle,fill=white] (c3pr2) at (10,-7) {$r_2$};

\draw[line width=5] (c1l1) -- (c1r1);
\draw (c1l1) -- (c1r2);
\draw (c1l2) -- (c1r1);
\draw[line width=5] (c1l2) -- (c1r2);

\draw[line width=5] (c2l1) -- (c2r1);
\draw (c2l1) -- (c2r2);
\draw (c2l2) -- (c2r1);
\draw (c2l2) -- (c2r2);

\draw (c3l1) -- (c3r1);
\draw (c3l1) -- (c3r2);
\draw (c3l2) -- (c3r1);
\draw (c3l2) -- (c3r2);

\draw (c2pl1) -- (c2pr1);
\draw[line width=5] (c2pl1) -- (c2pr2);
\draw[line width=5] (c2pl2) -- (c2pr1);

\draw[line width=5] (c3pl1) -- (c3pr1);
\draw[line width=5] (c3pl2) -- (c3pr2);
\end{tikzpicture} 
\caption{Three possible cycles induced by $h$. The thin edges correspond to $h_{\ell, r} = 1/3$ and thick edges correspond to $h_{\ell, r} = 2/3$.}\label{fig:cycle-breaking}
\end{figure}
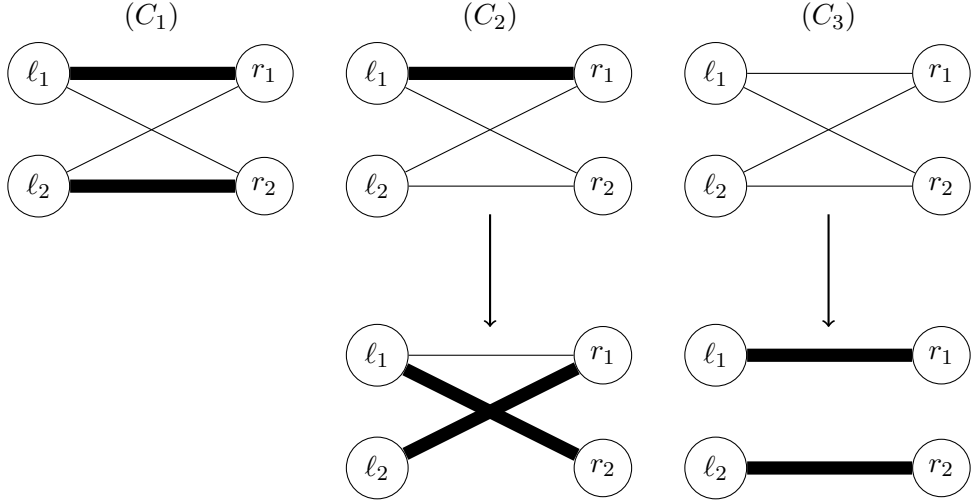

\textbf{Step 3} --- the first modification to $h$. Restrict the original type graph to a subgraph of edges $\{\ell, r\}$ such that $h_{\ell, r} > 0$. This graph is sparse --- each online node can have at most 3 neighbors. In Step 3, the goal is to break certain 4-cycles --- see Figure~\ref{fig:cycle-breaking} for details. Formally, this procedure is done by breaking all $(C_2)$-type cycles first. Then if there is a $(C_3)$-type cycle, break it. Return to trying to break $(C_2)$-cycles. This way you always try to break $(C_2)$ cycles first. This continues until all $(C_2)$ and $(C_3)$ cycles are broken.

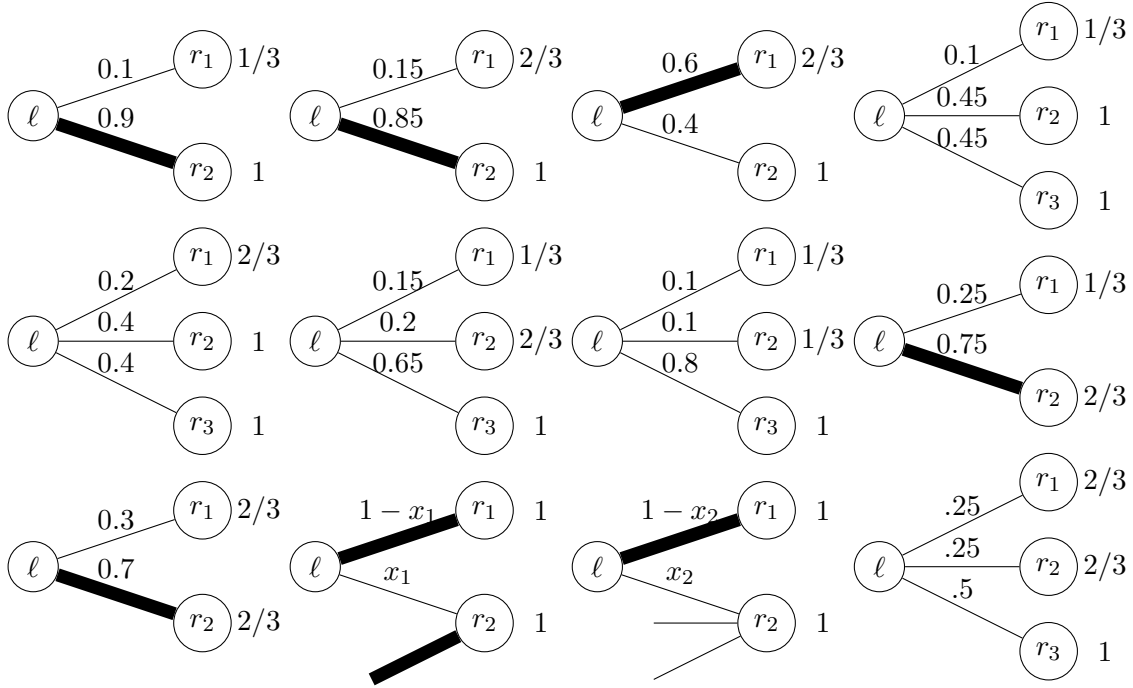
\begin{figure}[ht]
\centering

\begin{tikzpicture}[scale=0.75]


\node[draw=black,circle,fill=white] (c1l) at (0,0) {$\ell$};

\node[draw=black,circle,fill=white] (c1r1) at (3,1) {$r_1$};
\node[draw=black,circle,fill=white] (c1r2) at (3,-1) {$r_2$};

\node at (4,1) {$1/3$};
\node at (4,-1) {$1$};

\draw (c1l) -- node[above] {$0.1$} (c1r1) ;
\draw[line width=5] (c1l) -- node[above] {$0.9$} (c1r2);


\node[draw=black,circle,fill=white] (c2l) at (5,0) {$\ell$};

\node[draw=black,circle,fill=white] (c2r1) at (8,1) {$r_1$};
\node[draw=black,circle,fill=white] (c2r2) at (8,-1) {$r_2$};

\node at (9,1) {$2/3$};
\node at (9,-1) {$1$};

\draw (c2l) -- node[above] {$0.15$} (c2r1) ;
\draw[line width=5] (c2l) -- node[above] {$0.85$} (c2r2);


\node[draw=black,circle,fill=white] (c3l) at (10,0) {$\ell$};

\node[draw=black,circle,fill=white] (c3r1) at (13,1) {$r_1$};
\node[draw=black,circle,fill=white] (c3r2) at (13,-1) {$r_2$};

\node at (14,1) {$2/3$};
\node at (14,-1) {$1$};

\draw[line width=5]  (c3l) -- node[above] {$0.6$} (c3r1) ;
\draw (c3l) -- node[above] {$0.4$} (c3r2);


\node[draw=black,circle,fill=white] (c4l) at (15,0) {$\ell$};

\node[draw=black,circle,fill=white] (c4r1) at (18,1.5) {$r_1$};
\node[draw=black,circle,fill=white] (c4r2) at (18,0) {$r_2$};
\node[draw=black,circle,fill=white] (c4r3) at (18,-1.5) {$r_3$};

\node at (19,1.5) {$1/3$};
\node at (19,0) {$1$};
\node at (19,-1.5) {$1$};

\draw  (c4l) -- node[above] {$0.1$} (c4r1) ;
\draw (c4l) -- node[above] {$0.45$} (c4r2);
\draw (c4l) -- node[above] {$0.45$} (c4r3);


\node[draw=black,circle,fill=white] (c5l) at (0,-4) {$\ell$};

\node[draw=black,circle,fill=white] (c5r1) at (3,-2.5) {$r_1$};
\node[draw=black,circle,fill=white] (c5r2) at (3,-4) {$r_2$};
\node[draw=black,circle,fill=white] (c5r3) at (3,-5.5) {$r_3$};

\node at (4,-2.5) {$2/3$};
\node at (4,-4) {$1$};
\node at (4,-5.5) {$1$};

\draw  (c5l) -- node[above] {$0.2$} (c5r1) ;
\draw (c5l) -- node[above] {$0.4$} (c5r2);
\draw (c5l) -- node[above] {$0.4$} (c5r3);


\node[draw=black,circle,fill=white] (c6l) at (5,-4) {$\ell$};

\node[draw=black,circle,fill=white] (c6r1) at (8,-2.5) {$r_1$};
\node[draw=black,circle,fill=white] (c6r2) at (8,-4) {$r_2$};
\node[draw=black,circle,fill=white] (c6r3) at (8,-5.5) {$r_3$};

\node at (9,-2.5) {$1/3$};
\node at (9,-4) {$2/3$};
\node at (9,-5.5) {$1$};

\draw  (c6l) -- node[above] {$0.15$} (c6r1) ;
\draw (c6l) -- node[above] {$0.2$} (c6r2);
\draw (c6l) -- node[above] {$0.65$} (c6r3);


\node[draw=black,circle,fill=white] (c7l) at (10,-4) {$\ell$};

\node[draw=black,circle,fill=white] (c7r1) at (13,-2.5) {$r_1$};
\node[draw=black,circle,fill=white] (c7r2) at (13,-4) {$r_2$};
\node[draw=black,circle,fill=white] (c7r3) at (13,-5.5) {$r_3$};

\node at (14,-2.5) {$1/3$};
\node at (14,-4) {$1/3$};
\node at (14,-5.5) {$1$};

\draw  (c7l) -- node[above] {$0.1$} (c7r1) ;
\draw (c7l) -- node[above] {$0.1$} (c7r2);
\draw (c7l) -- node[above] {$0.8$} (c7r3);


\node[draw=black,circle,fill=white] (c8l) at (15,-4) {$\ell$};

\node[draw=black,circle,fill=white] (c8r1) at (18,-3) {$r_1$};
\node[draw=black,circle,fill=white] (c8r2) at (18,-5) {$r_2$};

\node at (19,-3) {$1/3$};
\node at (19,-5) {$2/3$};

\draw  (c8l) -- node[above] {$0.25$} (c8r1) ;
\draw[line width=5] (c8l) -- node[above] {$0.75$} (c8r2);


\node[draw=black,circle,fill=white] (c9l) at (0,-8) {$\ell$};

\node[draw=black,circle,fill=white] (c9r1) at (3,-7) {$r_1$};
\node[draw=black,circle,fill=white] (c9r2) at (3,-9) {$r_2$};

\node at (4,-7) {$2/3$};
\node at (4,-9) {$2/3$};

\draw  (c9l) -- node[above] {$0.3$} (c9r1) ;
\draw[line width=5] (c9l) -- node[above] {$0.7$} (c9r2);


\node[draw=black,circle,fill=white] (c11l) at (5,-8) {$\ell$};

\node[draw=black,circle,fill=white] (c11r1) at (8,-7) {$r_1$};
\node[draw=black,circle,fill=white] (c11r2) at (8,-9) {$r_2$};

\node at (9,-7) {$1$};
\node at (9,-9) {$1$};

\draw[line width=5]  (c11l) -- node[above] {$1-x_1$} (c11r1) ;
\draw (c11l) -- node[above] {$x_1$} (c11r2);
\draw[line width=5] (c11r2) -- (6, -10);


\node[draw=black,circle,fill=white] (c12l) at (10,-8) {$\ell$};

\node[draw=black,circle,fill=white] (c12r1) at (13,-7) {$r_1$};
\node[draw=black,circle,fill=white] (c12r2) at (13,-9) {$r_2$};

\node at (14,-7) {$1$};
\node at (14,-9) {$1$};

\draw[line width=5]  (c12l) -- node[above] {$1-x_2$} (c12r1) ;
\draw (c12l) -- node[above] {$x_2$} (c12r2);
\draw (c12r2) -- (11, -9);
\draw (c12r2) -- (11, -10);


\node[draw=black,circle,fill=white] (c10l) at (15,-8) {$\ell$};

\node[draw=black,circle,fill=white] (c10r1) at (18,-6.5) {$r_1$};
\node[draw=black,circle,fill=white] (c10r2) at (18,-8) {$r_2$};
\node[draw=black,circle,fill=white] (c10r3) at (18,-9.5) {$r_3$};

\node at (19,-6.5) {$2/3$};
\node at (19,-8) {$2/3$};
\node at (19,-9.5) {$1$};

\draw  (c10l) -- node[above] {$.25$} (c10r1) ;
\draw (c10l) -- node[above] {$.25$} (c10r2);
\draw  (c10l) -- node[above] {$.5$} (c10r3) ;

\end{tikzpicture} 
\caption{Three possible cycles induced by $h$. The thin edges correspond to $h_{\ell, r} = 1/3$ and thick edges correspond to $h_{\ell, r} = 2/3$. Numbers above edges correspond to new values of $h'$. The numbers next to $r$ nodes correspond to total values of $h$. The two magic numbers are $x_1 = 0.2744$ and $x_2 = 0.15877$.}\label{fig:second-mod}
\end{figure}

\textbf{Step 4} -- second modification to $h$. We call the result of this modification $h'$. This modification is presented in Figure~\ref{fig:second-mod}. In that figure, the numbers next to an offline node $r$ indicates the total value of $h$ at that node, i.e., $\sum_{\ell} h(\ell, r)$. Thin edges correspond to $h_{\ell, r} = 1/3$ and thick edges correspond to $h_{\ell, r} = 2/3$. The number above the edge corresponds to the newly assigned $h'$. For example, a thin edge with value $0.15$ above it means that $h_{\ell, r} = 1/3$ and after modification we have $h'(\ell, r) = 0.15$. Any edges not covered by one of the cases in the figure retain their old value of $h$.

Lastly, in \textbf{Step 5}, the distributions on lists are defined as follows. If $\ell$ has $1$ or $0$ neighbors in the sparse graph based on $h'$, then the distribution is fully supported on either the single-element list or the empty list, respectively. If $\ell$ has $2$ neighbors, say, $r_1$ and $r_2$ then the distribution is supported on two lists $(r_1, r_2)$ and $(r_2, r_1)$ with the probability of $r_1, r_2)$ being proportional to $h'(\ell, r_1)$. If the neighborhood of $\ell$ consists of three vertices, say, $r_1, r_2, r_3$, then the distribution is supported on all possible permutation of $(r_1, r_2, r_3)$, such that the probability that the list is $(r_i, r_j, r_k)$ is proportional to $\frac{h'(\ell, r_i) h'(\ell, r_j)}{h'(\ell, r_j) + h'(\ell, r_k)}$. Algorithm~\ref{alg:brubachetal} summarizes this procedure.

\begin{algorithm}
\caption{The known i.i.d. algorithm due to Brubach et al.~\cite{BrubachSSX16}.}\label{alg:brubachetal}
\begin{algorithmic}
\Procedure{BrubachEtAl}{$G=(L,R,E)$ -- type graph}
\State \Comment{Preprocessing stage:}
\State Solve LP~\eqref{eq:brubachetal}. Let $f^*$ denote an optimal solution. (\textbf{Step 1})
\State Scale $f^*$ multiplicatively to $3f^*$ and apply the rounding procedure of Gandhi et al.~\cite{GandhiKPS2006}. (\textbf{Step 2})
\State Set $h$ to be the scaled down (multiplicatively by $1/3$) rounded solution.
\State Apply the two modification steps to get $h'$. (\textbf{Steps 3 and 4})
\State Define the distributions on (sub)lists of neighbors. (\textbf{Step 5})
\State \Comment{Online stage:}
\State Run \textsc{RLA} with the above distribution.
\EndProcedure
\end{algorithmic}
\end{algorithm}

\subsubsection{Algorithms for Other Settings}

\noindent {\bf \large \textsc{Category-Advice}.} D\"urr \etal.  \cite{Durr2016} suggested a greedy-like algorithm that performs a second pass over the input called \textsc{Category-Advice}. The \textsc{Category-Advice} algorithm belongs to the class of category algorithms that were introduced in the work of D\"urr \etal. These algorithms are neither online nor known i.i.d. They can be viewed as conceptually simple offline algorithms, or online algorithms with advice (see~\cite{BoyarFKLM2016}), or as defining their own computational model.

A category algorithm starts with a permutation $\sigma$ of the offline nodes (e.g., given adversarially, or by an alphabetical order of names of the offline nodes). Instead of running \textsc{GreedyWithPermutation} directly with $\sigma$, the algorithm start by computing a category function $c: R \rightarrow \mathbb{Z}$. The algorithm updates $\sigma$ to $\sigma_c$ as follows: $\sigma_c$ is the unique permutation satisfying that for all $v_1, v_2 \in R$, we have $\sigma_c(v_1) < \sigma_c(v_2)$ if and only if $c(v_1) < c(v_2)$ or ($c(v_1) = c(v_2)$ and $\sigma(v_1) < \sigma(v_2)$). Then \textsc{GreedyWithPermutation} is performed with $\sigma_c$ as the permutation of the offline nodes. In other words, a category algorithm partitions the offline nodes into $|\Ima(c)|$ categories and specifies the ranking of the categories, the ranking within the category is induced by the initial permutation $\sigma$.

The \textsc{Category-Advice} algorithm starts with $\sigma$, and in the first pass, runs \textsc{GreedyWithPermutation} with $\sigma$. Let $M$ be the matching obtained in the first pass. The category function $c: R \rightarrow [2]$ is defined as follows: $c(v) = 1$ if $v$ does not participate in $M$ and $c(v) = 2$ otherwise. In the second pass, the \textsc{Category-Advice} algorithm runs \textsc{GreedyWithPermutation} with $\sigma_c$. The output of the second run of \textsc{GreedyWithPermutation} is declared as the output of the \textsc{Category-Advice} algorithm. In other words, in the second pass the algorithm gives preference to those vertices that were {\bf not} matched in the first pass. Algorithm~\ref{alg:category-advice} shows the pseudocode.

\begin{algorithm}
\caption{The \textsc{Category-Advice} algorithm of D\"urr et al.~\cite{Durr2016}.}\label{alg:category-advice}
\begin{algorithmic}
\Procedure{Category-Advice}{$G=(L,R,E), \sigma : R \rightarrow R$}
\State Set $M = \textsc{GreedyWithPermutation}(G,\sigma)$
\State Define $c : R \rightarrow [2]$ by $c(v) = 1$ if $M(v) = \bot$ and $c(v) = 2$ otherwise.
\State Define $\sigma_c$ as stated in the main text.
\State Return $\textsc{GreedyWithPermutation}(G,\sigma_c)$
\EndProcedure
\end{algorithmic}
\end{algorithm}

\noindent {\bf \large \textsc{3-Pass}.} The algorithm of D\"urr et al. was extended to multiple passes in \cite{BorodinPS2018}. In this paper, we shall only consider the generalization of the algorithm to 3 passes, which we call \textsc{3-Pass}. In the first two passes, the algorithm behaves the same way as \textsc{Category-Advice}. The generalization is quite natural: in the third pass, the algorithm prefers to match an incoming node to an offline node that was not matched in the 1st or 2nd pass. If there is no such node available, then \textsc{3-Pass} prefers to match an incoming node to an offline node that was not matched in the 1st pass. If there is no such offline node, then \textsc{3-Pass} matches an incoming node to the first (according to the original fixed ordering) available offline node.

\subsection{Conversion to Greedy}
\label{sec:greedy-conversion}

As mentioned in Remark~\ref{rem:greedy}, all of the complicated known i.i.d. algorithms from the previous section are presented in the corresponding papers as non-greedy to simplify the analysis. For example, suppose that $u$ is an online node of type $\ell$. Moreover, assume that it is the third arrival of type $\ell$ and consider the behavior of \textsc{FeldmanEtAl}. Regardless of how many neighbors of $u$ are available, \textsc{FeldmanEtAl} is not going to match $u$ (\textsc{FeldmanEtAl} only attempts to match first and second arrivals of a given type). A greedy algorithm would match $u$ if it had at least one available neighbor. Similar considerations hold for the rest of the algorithms in that section. Thus, vanilla versions of these algorithms immediately forgo a constant fraction of possible matches in order to simplify the analysis and optimize for the worst-case. Clearly, there are type graphs (e.g., the complete type graph), on which any greedy algorithm would be able to find a nearly-perfect matching. On such graphs, the complicated algorithms would be vastly outperformed by any greedy algorithm.

Fortunately, there is a simple idea  to turn all of these algorithms into greedy ones while preserving their worst-case guarantees. The idea is just to run a greedy algorithm, and if there are several available neighbors, break ties by using the suggestions of these algorithms. For instance, if we wish to convert \textsc{FeldmanEtAl} into a greedy algorithm, it would work as follows. Let $u$ be an online node of type $\ell$. If $u$ is the first arrival of type $\ell$, try to match $u$ to its blue neighbor. If blue neighbor is not available, match it to the first available neighbor. If $u$ is the second arrival of type $\ell$, try to match $u$ to its red neighbor. If red neighbor is not available, match it to the first available neighbor. If $u$ is the third (and onward) arrival of type $\ell$, match $u$ to the first available neighbor. It is a simple exercise to show that this modification always constructs a matching that is at least as large as that found by \textsc{FeldmanEtAl}. Moreover, this modification is easy to implement and does not seem to have a significant affect on the runtime. Similar modifications can be made to other algorithms. In our experiments, we report performance of both greedy and non-greedy versions of known i.i.d. algorithms.
\subsection{Notes on Implementation}
\label{sec:implmentation-notes}

We used the adjacency list representation of graphs for all of the above algorithms. Compared to adjacency matrix representation, this allowed for significant speedup on sparse graphs.

The max flow problems in \textsc{FeldmanEtAl} and \textsc{BahmaniKapralov} are solved via straightforward implementations of Edmonds-Karp max flow algorithm. The same algorithm is used to obtain an optimal maximum matching. 

For \textsc{ManshadiEtAl}, we estimate a fractional optimal solution by running Edmonds-Karp algorithm initialized with a greedy solution (for speed) on 100 samples generated for a given type graph.

The linear program~\eqref{eq:jailletlu} in \textsc{JailletLu} can be formulated as a max flow problem with integral capacities. This is done by rescaling constraints by a multiple of 3, and constructing the following flow network. Add a source $s$ and a sink $t$, connect $s$ to each $r \in R$ via edges of capacity $3$, connect each $\ell \in L$ to $t$ via edges of capacity $3$, orient edges of $G$ from $R$ to $L$ and assign capacity $2$ to them. In our implementation, we use Edmonds-Karp to solve this max flow problem via an integral flow. Then $f^*$ can be obtained by scaling the max flow by a multiple of $1/3$.

We solve the linear program~\eqref{eq:brubachetal} in \textsc{BrubachEtAl} using the simplex method in GNU Linear Programming Kit (GLPK)~\cite{GLPK}.

The actual code is freely available at~\cite{BMCODE}.

\section{Experimental Setup}
\label{sec:experimental_setup}
All our experiments were performed on a personal laptop with Intel Core i5-7300HQ processor clocked at 2.5 Ghz. The laptop had 8 GB 2400 Mhz DDR4 of RAM and 256GB M.2 SSD. The laptop was running Windows 10 64-bit Home edition. All algorithms under consideration were coded in C++ and compiled with Microsoft Visual Studio Community 2017 version 15.5.7. The code was compiled for the 64-bit target architecture with an optimization flag O2. The implementation is single-threaded, so all algorithm runs were performed on a single core.

In the rest of this section we describe our benchmarks for online bipartite matching algorithms under the known i.i.d. input model with integral types. Our benchmarks can naturally be split into three categories; namely, parameterized families of graphs, stand-alone graphs, and bipartite graphs derived from real-world graphs (which we will call ``real-world graphs'' for short). Graphs in these categories refer to \emph{type graphs} with the understanding that \emph{instance graphs} corresponding to a particular type graph from the benchmark will be obtained by sampling $n$ online nodes uniformly at random from all possible types, i.i.d.

Families of graphs are obtained by either a random or a deterministic process that has a natural parameter. For example, this parameter could be a proxy for edge density of a graph. For families of graphs, we will be interested in the performance of the algorithms as a function of the given parameter.

We call a graph stand-alone if it is obtained either by a random or a deterministic process, but there are no associated parameters. For example, worst-case graphs for online algorithms. Although all graphs can be parameterized by the size of the graph, we are interested in the asymptotic behavior of the algorithms on large graphs, so we typically take stand-alone graphs of largest size that can be solved in reasonable time by all algorithms under consideration. Note that stand-alone graphs are not necessarily fixed and can still be the result of a random process.

Stand-alone graphs \textsc{FewG, ManyG, Rope, Hexa, Zipf} are taken from Cherkassky et al.~\cite{CherkasskyGMSS1998}, where these graphs were used to measure the performance of various offline algorithms for bipartite matching. Our implementation of the generating procedures for these graphs does not perfectly match the code accompanying the paper~\cite{CherkasskyGMSS1998}, because their code is designed for more general families of graphs. Instead, our implementation follows the descriptions in the paper~\cite{CherkasskyGMSS1998} itself, where parameters are often fixed to certain values that simplify the generating process.

We also consider a number of graphs that are publicly available from online repositories. 

Most of our synthetically generated instances are bipartite. A few of our synthetically generated instances, as well as all real-world instances are non-bipartite. In case of a non-bipartite graph, we use one of the following two ways of creating a bipartite graph out of a non-bipartite graph. Let $G = (V, E)$ be a given graph that is not necessarily bipartite. We call the first way of creating a bipartite graph out of $G$ \emph{the duplicating method}. The idea is to duplicate the vertex set $V$. Let $L = V$ be the first copy of $V$ and $R = V$ be the second copy. Put an edge between $\ell \in L$ and $r \in R$ if and only if $\{\ell, r\} \in E$. We call the second way of creating a bipartite graph out of $G$ \emph{the random balanced partition method}. In this method, we partition $V$ randomly into two blocks $L$ and $R$, such that $|L| = \lfloor |V|/2 \rfloor$ and $|R| = \lceil |V|/2 \rceil$. We keep only those edges that connect two vertices from different partitions. Solving the matching problem on a graph obtained from the random balanced partition method applied to social network graphs has a natural interpretation. This corresponds to dividing the whole population into two groups and pairing up as many ``friends'' (``co-authors'', ``co-stars'', etc.) from the two groups as possible.


\subsection{Families of Graphs}
\label{subsec:families}

\noindent {\bf \large \textsc{Erd\H{o}s-R\'{e}nyi Graphs}.} A graph of this family is denoted by $G_{n,n,p}$. We have that $|L|=|R|=n$ and for each $\ell \in L$ and $r \in R$ an edge $\{\ell, r\}$ is included in $G$ with probability $p$ independently. We consider $p$ to be of the form $c/n$ and $c$ is the parameter defining this family of graphs.
\\\\
\noindent {\bf \large \textsc{Random Regular on the Left (Right) Graphs}.} We say that a graph $G$ is $d$-regular on the left (right) if the degree of every vertex in $L$ (in $R$) is the same and equal to $d$. To generate a random graph that is $d$-regular on the left, for each $\ell$ we sample a uniformly random subset of $d$ vertices from $R$ and declare them to be neighbors of $\ell$. The samples for different $\ell$ are independent. The procedure to generate $d$-regular graphs on the right is analogous. These families of graphs are parameterized by $d$. 
\\\\
\noindent {\bf \large \textsc{Molloy-Reed}.} Molloy and Reed~\cite{MolloyR1995} gave a procedure to generate a graph with a given degree  distribution $p$. We describe the procedure for non-bipartite graphs. To generate a graph on $n$ nodes, for each node $u$ sample its degree from $p$. Initially, degree $d$ of $u$ corresponds to $d$ non-paired ends of edges. The idea is to choose randomly two such ends of edges and connect them together -- this forms an edge and decreases the number of non-paired edges by 1 for each of the 2 participating vertices. While there are vertices with non-paired edges, pick two such vertices at random and pair up one end of an edge from the first vertex with one end of an edge from the second vertex. There are a couple of problems with this procedure as stated. First of all, if the sum of all degrees is odd, this procedure will leave one end of an edge non-paired. This is fixed by modifying the first step --- after sampling degrees of vertices and before pairing up any ends of edges. While the total degree is odd, pick a random vertex and resample its degree. The second problem is that this procedure does not necessarily generate a \emph{simple graph} -- i.e., there might be self-loops and duplicate (parallel) edges. To address this issue, when pairing up edges, we perform 100 random samples of pairs of vertices to try and find ends of edges that do not result in self-loops or parallel edges. If all of these trials fail, then we add the self-loop or the parallel edge of the last trial. At the end of the procedure we obtain the graph by removing all self-loops and parallel edges.

While Erd\H{o}s-R\'{e}nyi model is natural, it does not seem to model many real-life scenarios, such as social networks. It has been long observed that degree distributions of many social networks (e.g., Facebook, Twitter, movie actor databases, researcher co-authorship databases, etc.) are not binomial, but rather seem to have heavy tails. Thus, they are more accurately modeled by power law distributions. Newman et al.~\cite{NewmanWS2002} describe a particular family of distributions that combined with Molloy-Reed procedure results in a fairly accurate model of many social networks. This family  of distributions is called a power law distribution with exponential cutoff. This distribution has two parameters -- $\tau$, which is called the exponent, and $\kappa$, which is called the cutoff. The idea is that for small values of $d$ the probability of a node having degree $d$ should be modeled by $x^{-\tau}$ (the power law part), but for $d > \kappa$ the probability should be dropping off exponentially (the exponential cutoff part). Formally, it is defined as follows. Let $p_d$ denote the probability of our random variable having value $d$, then we have:
\[p_d = \left\{
\begin{array}{ll}
0 & \text{if $d = 0$} \\
c x^{-\tau} e^{-d/\kappa} & \text{if $d > 0$},\\
\end{array}
\right.\]
where $c$ is the normalizing constant. The Molloy-Reed procedure on a power law distribution with exponential cutoff, followed by the random balanced partition method defines a family of type graphs that is parameterized by $\tau$ and $\kappa$.
\\\\
\noindent {\bf \large \textsc{Preferential Attachment Bigraphs}.}
We also consider the following natural modification of the preferential attachment model that immediately produces bipartite graphs without having to use the random balanced partition method. We refer to this model as the \emph{preferential attachment bigraph} model. To generate a bigraph in this model, start with $n$ offline nodes $R$ and introduce online nodes $L$ one at a time. The model has a single parameter $c$ which is the average degree of an online node. When a new online node $i \in L$ arrives, sample $Z_i \sim \Bin(n, c/n)$ to decide on a number of its offline neighbors. Let $d_j$ denote the current degree of an offline node $j \in R$. Define a probability distribution $\mu$ on offline nodes such that $\mu(j) = \frac{1+d_j}{n+\sum_{t \in R} d_t}$. Sample RHS nodes from $\mu$ i.i.d. repeatedly until $Z_i$ \emph{unique} offline nodes are generated. These offline nodes define the neighborhood of the current online node $i$. Update the $d_j$ and continue.
\\\\

\subsection{Stand-Alone Graphs}
\noindent {\bf \large \textsc{Upper-Triangular (UT)}.}This graph is the fixed graph defined by an upper-triangular adjacency matrix with the columns representing online nodes that arrive from right to left. This is known to be the worst case example for \textsc{Ranking} in the adversarial online model~\cite{KarpVV90}.
\begin{figure}[H]
	\centering
	
	\begin{tikzpicture}[scale=0.75]
	
	\node at (-6,1) {$(L)$};
	\node at (-2,1) {$(R)$};
	
	\node[draw=black,circle,fill=black] (c1l1) at (-6,0) {$ $};
	\node[draw=black,circle,fill=black] (c1l2) at (-6,-1) {$ $};
	\node[draw=black,circle,fill=black] (c1l3) at (-6,-2) {$ $};
	\node[draw=none] (c1l4) at (-6,-4) {$\vdots$};
	\node[draw=black,circle,fill=black] (c1l5) at (-6,-6) {$ $};

	\node[draw=black,circle,fill=black] (c1r1) at (-2,0) {$ $};
	\node[draw=black,circle,fill=black] (c1r2) at (-2,-1) {$ $};
	\node[draw=black,circle,fill=black] (c1r3) at (-2,-2) {$ $};
	\node[draw=none] (c1r4) at (-2,-4) {$\vdots$};
	\node[draw=black,circle,fill=black] (c1r5) at (-2,-6) {$ $};

	\draw (c1l1) -- (c1r1);
	\draw (c1l1) -- (c1r2);
	\draw (c1l1) -- (c1r3);
	\draw (c1l1) -- (c1r5);

	\draw (c1l2) -- (c1r2);
	\draw (c1l2) -- (c1r3);
	\draw (c1l2) -- (c1r5);

    \draw (c1l3) -- (c1r3);
	\draw (c1l3) -- (c1r5);
	
	\draw (c1l5) -- (c1r5);


	\end{tikzpicture} 
	\caption{Upper Triangular graph}\label{fig:ut_graph}
\end{figure}
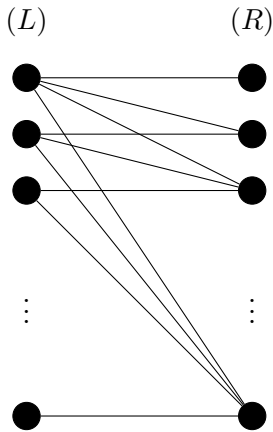

\noindent {\bf \large \textsc{Manshadi-Hard (MH)}.} In \cite{Manshadi2011}, Manshadi et al. present a type graph for which no online algorithm can achieve an expected competitive ratio better than $1-\frac{1}{e^{2}} \approx 0.86$. Let $G(L,R,E)$ be the type graph where $L=L_{1}\cup L_{2}$, $|L_{1}|=|R|=n$ and $L_{2}=n/e$. There is a perfect matching between the vertices of $L_{1}$ and $R$ and a complete bipartite graph between $L_{2}$ and $R$. Each type has arrival rate 1 and there are $|L|=n(1+1/e)$ online i.i.d. draws.
\begin{figure}[H]
	\centering
	
	\begin{tikzpicture}[scale=0.5]
	
	\node at (-6,1) {$(L)$};
	\node at (0,1) {$(R)$};
	
	\node[draw=black,circle,fill=black] (c1l1) at (-6,0) {$ $};
	\node[draw=black,circle,fill=black] (c1l2) at (-6,-1) {$ $};
	\node[draw=black,circle,fill=black] (c1l3) at (-6,-2) {$ $};
	\node[draw=none] (c1l4) at (-6,-3) {$\vdots$};
	\node[draw=black,circle,fill=black] (c1l5) at (-6,-4) {$ $};
	
	\node[draw=black,circle,fill=black] (c1l6) at (-6,-6) {$ $};
	\node[draw=black,circle,fill=black] (c1l7) at (-6,-7) {$ $};
	\node[draw=none] (c1l8) at (-6,-8) {$\vdots$};
	\node[draw=black,circle,fill=black] (c1l9) at (-6,-9) {$ $};

	\node[draw=black,circle,fill=black] (c1r1) at (0,0) {$ $};
	\node[draw=black,circle,fill=black] (c1r2) at (0,-1) {$ $};
	\node[draw=black,circle,fill=black] (c1r3) at (0,-2) {$ $};
	\node[draw=none] (c1r4) at (0,-3) {$\vdots$};
	\node[draw=black,circle,fill=black] (c1r5) at (0,-4) {$ $};

	\draw (c1l1) -- (c1r1);
	\draw (c1l2) -- (c1r2);
	\draw (c1l3) -- (c1r3);
	\draw (c1l5) -- (c1r5);
	
	\draw (c1l6) -- (c1r1);
	\draw (c1l6) -- (c1r2);
	\draw (c1l6) -- (c1r3);
	\draw (c1l6) -- (c1r5);
	\draw (c1l7) -- (c1r1);
	\draw (c1l7) -- (c1r2);
	\draw (c1l7) -- (c1r3);
	\draw (c1l7) -- (c1r5);
	\draw (c1l9) -- (c1r1);
	\draw (c1l9) -- (c1r2);
	\draw (c1l9) -- (c1r3);
	\draw (c1l9) -- (c1r5);

	\draw [decorate,decoration={brace,amplitude=10pt,mirror},xshift=-9pt,yshift=-5pt]
	(-6,0.5) -- (-6,-4.2) node [black,midway,xshift=-1.2cm] 
	{\footnotesize $|L_1|=n$};
	
	\draw [decorate,decoration={brace,amplitude=10pt,mirror},xshift=-9pt,yshift=-5pt]
	(-6,-5.5) -- (-6,-9.2) node [black,midway,xshift=-1.2cm] 
	{\footnotesize $|L_2|=n/e$};

	\end{tikzpicture} 
	\caption{Manshadi-Hard graph}\label{fig:mh_graph}
\end{figure}
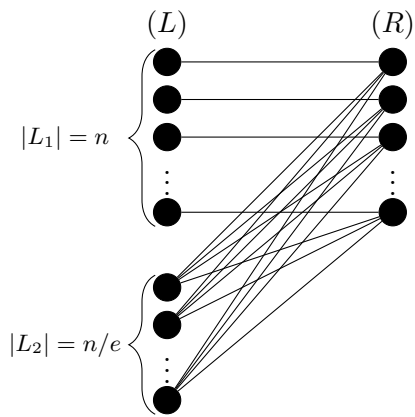

\noindent {\bf \large \textsc{Feldman-Hard (FH)}.} Feldman et al.~\cite{FeldmanMMM09} present a family of graphs that is the worst case for their algorithm, proving that their analysis of the competitive ratio of their algorithm is tight. $R$ is partitioned into 4 blocks: $K, U, V,$ and $W$, each of size $n/4$. Similarly, $L$ is partitioned into 4 blocks: $I, X, Y,$ and $Z$, each of size $n/4$. We use a lower-case letter to refer to an element in the given block, e.g., elements of $U$ are denoted by $u_i$, where $i \in [n/4]$. The edge set consists of a 6-cycle $(u_{i},x_{i},v_{i},y_{i},w_{i},z_{i},u_{i})$ for $i\in [1,\frac{n}{4}]$, a complete bipartite between $K$ and $X$, and a complete bipartite graph between $I$ and $W$.
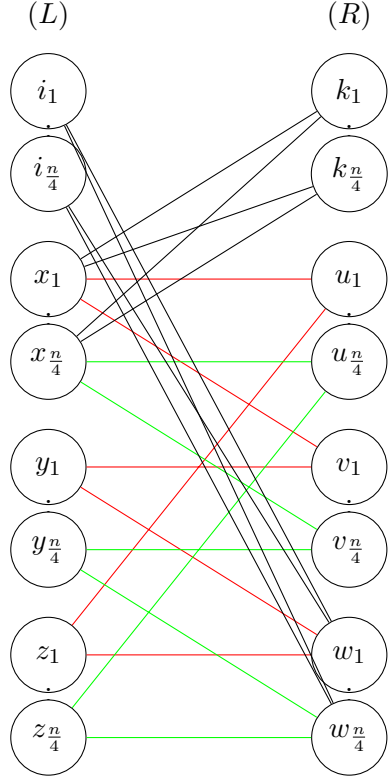
\begin{figure}[H]
	\centering
	
	\begin{tikzpicture}[scale=0.5,state/.style={circle, draw, minimum size=1cm}]
	
	\node at (-6,2) {$(L)$};
	\node at (2,2) {$(R)$};
	
	\node[state,draw=black,circle,fill=white] (i1) at (-6,0) {$i_1$};
	\node[draw=none] (i2) at (-6,-1) {$\vdots$};
	\node[state,draw=black,circle,fill=white] (in) at (-6,-2.2) {$i_{\frac{n}{4}}$};
	
	\node[state,draw=black,circle,fill=white] (x1) at (-6,-5) {$x_1$};
	\node[draw=none] (x2) at (-6,-6) {$\vdots$};
	\node[state,draw=black,circle,fill=white] (xn) at (-6,-7.2) {$x_{\frac{n}{4}}$};
	
	\node[state,draw=black,circle,fill=white] (y1) at (-6,-10) {$y_1$};
	\node[draw=none] (y2) at (-6,-11) {$\vdots$};
	\node[state,draw=black,circle,fill=white] (yn) at (-6,-12.2) {$y_{\frac{n}{4}}$};
	
	\node[state,draw=black,circle,fill=white] (z1) at (-6,-15) {$z_1$};
	\node[state,draw=none] (z2) at (-6,-16) {$\vdots$};
	\node[state,draw=black,circle,fill=white] (zn) at (-6,-17.2) {$z_{\frac{n}{4}}$};

	\node[state,draw=black,circle,fill=white] (k1) at (2,0) {$k_1$};
	\node[draw=none] (k2) at (2,-1) {$\vdots$};
	\node[state,draw=black,circle,fill=white] (kn) at (2,-2.2) {$k_{\frac{n}{4}}$};
	
	\node[state,draw=black,circle,fill=white] (u1) at (2,-5) {$u_1$};
	\node[draw=none] (u2) at (2,-6) {$\vdots$};
	\node[state,draw=black,circle,fill=white] (un) at (2,-7.2) {$u_{\frac{n}{4}}$};
	
	\node[state,draw=black,circle,fill=white] (v1) at (2,-10) {$v_1$};
	\node[draw=none] (v2) at (2,-11) {$\vdots$};
	\node[state,draw=black,circle,fill=white] (vn) at (2,-12.2) {$v_{\frac{n}{4}}$};
	
	\node[state,draw=black,circle,fill=white] (w1) at (2,-15) {$w_1$};
	\node[state,draw=none] (w2) at (2,-16) {$\vdots$};
	\node[state,draw=black,circle,fill=white] (wn) at (2,-17.2) {$w_{\frac{n}{4}}$};

	\draw[red] (u1) -- (x1) -- (v1) -- (y1) -- (w1) -- (z1) -- (u1);
	\draw[green] (un) -- (xn) -- (vn) -- (yn) -- (wn) -- (zn) -- (un);
	
	\draw (i1) -- (w1);
	\draw (i1) -- (wn);
	\draw (in) -- (w1);
	\draw (in) -- (wn);
	
	\draw (k1) -- (x1);
	\draw (k1) -- (xn);
	\draw (kn) -- (x1);
	\draw (kn) -- (xn);

	\end{tikzpicture} 
	\caption{Feldman-Hard graph. Depicted with red and green edges are the cycles $(u_{1},x_{1},v_{1},y_{1},w_{1},z_{1},u_{1})$ and $(u_{n/4},x_{n/4},v_{n/4},y_{n/4},w_{n/4},z_{n/4},u_{n/4})$ respectively. The remaining edges form two complete bipartite graphs}\label{fig:fh_graph}
\end{figure}

\noindent {\bf \large \textsc{FewG and ManyG}.} To construct these bipartite graphs, the vertices in $L$ are randomly permuted and then $L$ and $R$ are partitioned into $k$ groups of equal size. Each vertex of the $i\mbox{-}$th group of $L$ is assigned $Y$ random neighbors from the $(i-1)\mbox{-}$th through $(i+1)\mbox{-}$th group of $R$ (with wrap around). To be consistent with previous literature (\cite{CherkasskyGMSS1998}) $Y$ is set to be binomially distributed with $\mathbb{E}(Y)=5$, and we consider the two cases of $k=32$ {\bf \large \textsc{(FewG)}} and $k=256$ {\bf \large \textsc{(ManyG)}}.
\begin{figure}[H]
	\centering
	
	\begin{tikzpicture}[scale=0.5,state/.style={rectangle, draw, minimum size=1.1cm}]
	
	\node at (-6,2) {$(L)$};
	\node at (0,2) {$(R)$};
	
	\node[state,draw=black,rectangle,fill=white](l0) at (-6,0) {$L_0$};
	\node[draw=none] (ldots1) at (-6,-3) {$\vdots$};
	\node[state,draw=black,rectangle,fill=white] (li) at (-6,-6) {$L_i$};
	\node[draw=none] (ldots2) at (-6,-8) {$\vdots$};
	\node[state,draw=black,rectangle,fill=white] (lk1) at (-6,-11) {$L_{k-1}$};
	
	\node[state,draw=black,rectangle,fill=white](r0) at (0,0) {$R_0$};
	\node[draw=none] (i2) at (0,-1.7) {$\vdots$};
	\node[state,draw=black,rectangle,fill=white] (ri0) at (0,-4) {$R_{i-1}$};
	\node[state,draw=black,rectangle,fill=white] (ri) at (0,-6) {$R_{i}$};
	\node[state,draw=black,rectangle,fill=white] (ri1) at (0,-8) {$R_{i+1}$};
	\node[draw=none] (rdots) at (0,-9.2) {$\vdots$};
	\node[state,draw=black,rectangle,fill=white] (rk) at (0,-11) {$R_{k-1}$};

	\draw[line width=2] (li) -- (ri0);
	\draw[line width=2] (li) -- (ri);
	\draw[line width=2] (li) -- (ri1);


	\end{tikzpicture} 
	\caption{Graph FewG and ManyG. For $i=0,\dots ,k-1$, vertices in $L_{i}$ are assigned random neighbors from groups $R_{i-1}$ to $R_{i+1}$.}\label{fig:fewmanyg_graph}
\end{figure}
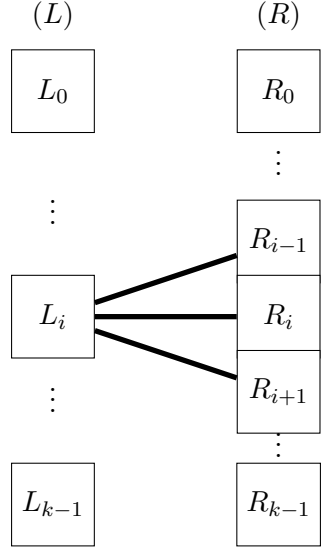

\noindent {\bf \large \textsc{rope}.} For this graph, vertices in $L$ and $R$ are grouped into $t=n/d$ groups of size $d$, denoted $L_{0}\dots L_{t-1}$ and $R_{0}\dots R_{t-1}$. Block $i$ on one side is connected to block $i+1$ on the other side, for $i = 0\dots t-2$; block $L_{t-1}$ is connected to block $R_{t-1}$. Thus the graph is a ``rope'' that zigzags between the two sides of the graph, first up and then down. Consecutive pairs of blocks along the rope are connected alternately by perfect matchings and random bipartite graphs of average degree $d-1$, beginning and ending with perfect matchings. As in \cite{CherkasskyGMSS1998}, we fix $d=6$.
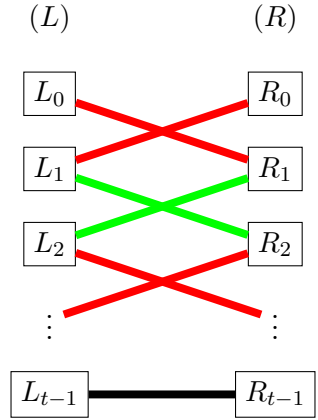
\begin{figure}[H]
	\centering
	
	\begin{tikzpicture}[scale=0.5]
	
	\node at (-6,2) {$(L)$};
	\node at (0,2) {$(R)$};
	
	\node[draw=black,rectangle,fill=white](l0) at (-6,0) {$L_0$};
	\node[draw=black,rectangle,fill=white] (l1) at (-6,-2) {$L_1$};
	\node[draw=black,rectangle,fill=white] (l2) at (-6,-4) {$L_2$};
	\node[draw=none] (ldots) at (-6,-6) {$\vdots$};
	\node[draw=black,rectangle,fill=white] (lt1) at (-6,-8) {$L_{t-1}$};
	
	\node[draw=black,rectangle,fill=white](r0) at (0,0) {$R_0$};
	\node[draw=black,rectangle,fill=white] (r1) at (0,-2) {$R_1$};
	\node[draw=black,rectangle,fill=white] (r2) at (0,-4) {$R_2$};
	\node[draw=none] (rdots) at (0,-6) {$\vdots$};
	\node[draw=black,rectangle,fill=white] (rt1) at (0,-8) {$R_{t-1}$};

	\draw[red,line width=3] (l0) -- (r1);
	\draw[green,line width=3] (l1) -- (r2);
	\draw[red,line width=3] (l2) -- (rdots);
	
	\draw[red,line width=3] (r0) -- (l1);
	\draw[green,line width=3] (r1) -- (l2);
	\draw[red,line width=3] (r2) -- (ldots);
	
	\draw[line width=3] (lt1) -- (rt1);
	
	\end{tikzpicture} 
	\caption{Graph Rope. Red edges correspond to perfect matchings and green edges correspond to random bipartite graphs. Edge $(L_{t-1},R_{t-1})$ can be either red or green.}\label{fig:rope_graph}
\end{figure}

\noindent {\bf \large \textsc{hexa}.} In these graphs, each side is partitioned into $\sqrt{n}$ blocks of size $\sqrt{n}$ each. A random bipartite hexagon is added between block $i$ on one side and block $j$ on the other side for all $i,j \in [\sqrt{n}]$. This results in the average degree of each vertex being 6. The random hexagon is generated by the following procedure. Pick 3 random nodes on each side (inside the corresponding blocks), say, $\ell_1, \ell_2, \ell_3$ and $r_1, r_2, r_3$. Sample two random permutations $\pi, \sigma : [3] \rightarrow [3]$. Add the following cycle to the graph $(\ell_{\pi(1)}, r_{\sigma(1)}, \ell_{\pi(2)}, r_{\sigma(2)}, \ell_{\pi(3)}, r_{\sigma(3)}, \ell_{\pi(1)})$.
\begin{figure}[H]
	\centering
	
	\begin{tikzpicture}[scale=0.5,state/.style={rectangle, draw, minimum size=1.2cm}]
	
	\node at (-6,2) {$(L)$};
	\node at (0,2) {$(R)$};
	
	\node[state,draw=black,rectangle,fill=white](l0) at (-6,0) {$L_0$};
	\node[state,draw=black,rectangle,fill=white] (l1) at (-6,-3) {$L_1$};
	\node[state,draw=black,rectangle,fill=white] (l2) at (-6,-6) {$L_2$};
	\node[draw=none] (ldots2) at (-6,-8) {$\vdots$};
	\node[state,draw=black,rectangle,fill=white] (lsqn) at (-6,-10) {$L_{\sqrt{n}-1}$};
	
	\node[state,draw=black,rectangle,fill=white](r0) at (0,0) {$R_0$};
	\node[state,draw=black,rectangle,fill=white] (r1) at (0,-3) {$R_1$};
	\node[state,draw=black,rectangle,fill=white] (r2) at (0,-6) {$R_2$};
	\node[draw=none] (rdots2) at (0,-8) {$\vdots$};
	\node[state,draw=black,rectangle,fill=white] (rsqn) at (0,-10) {$R_{\sqrt{n}-1}$};

	\draw[line width=2] (l0) -- (r0);
	\draw[line width=2] (l0) -- (r1);
	\draw[line width=2] (l0) -- (r2);
	\draw[line width=2] (l0) -- (rsqn);
	
	\draw[line width=2] (l1) -- (r0);
	\draw[line width=2] (l1) -- (r1);
	\draw[line width=2] (l1) -- (r2);
	\draw[line width=2] (l1) -- (rsqn);
	
	\draw[line width=2] (l2) -- (r0);
	\draw[line width=2] (l2) -- (r1);
	\draw[line width=2] (l2) -- (r2);
	\draw[line width=2] (l2) -- (rsqn);
	
	\draw[line width=2] (lsqn) -- (r0);
	\draw[line width=2] (lsqn) -- (r1);
	\draw[line width=2] (lsqn) -- (r2);
	\draw[line width=2] (lsqn) -- (rsqn);

	\end{tikzpicture} 
	\caption{Graph Hexa. Connecting groups form a complete bipartite graph. Each edge depicts a random hexagon between the corresponding groups.}\label{fig:hexa_graph}
\end{figure}
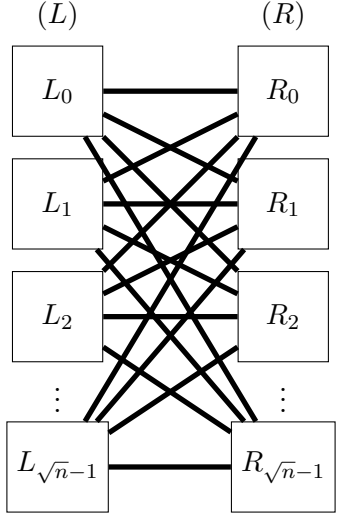

\noindent {\bf \large \textsc{zipf}.} In these bipartite graphs, we have $|L|=|R|=n$ and an edge between nodes 
$\ell_i \in L$ and $r_j \in R$ exists with probability roughly proportional to $1/ij$. More precisely, the probability is $\Pr(\ell_i \sim r_j)=\min\left(\frac{n\cdot d}{\log^{2}n}\cdot\frac{1}{i\cdot j},1 \right)$ with $d=6$ and $i, j \in [n]$. This results in graphs that are denser around vertices with smaller indices.\\\\

\subsection{Real-World Data} 
To perform experiments on real datasets, we used some publicly available graphs from the Network Data Repository \cite{NetworkRepo}. In the experiments we used both the duplicating method and the random balanced partition method of bipartite transformations.

The \textbf{socfb} datasets are social frienship networks extracted from Facebook. Nodes are users and edges represent friendship ties. The \textbf{bio-CE} datasets correspond to biological datasets representing links by similar phylogenetic profiles and gene neighbourhoods of bacterial and archaeal orthologs. We also used two \textbf{econ} datasets that model US economic transactions in 1972 by connecting commodities to commodities and industries. These datasets, along with various properties of the corresponding graphs, can be found in the following links\footnote{Accessed 2018-05-25}:
\begin{itemize}
    \item http://networkrepository.com/socfb-Caltech36.php
    \item http://networkrepository.com/socfb-Reed98.php
    \item http://networkrepository.com/bio-CE-GN.php
    \item http://networkrepository.com/bio-CE-PG.php
    \item http://networkrepository.com/econ-mbeaflw.php
    \item http://networkrepository.com/econ-beause.php
\end{itemize}

\section{Experimental Results}
\label{sec:experimental_results}
In this section, we present results of our experiments and provide some comments about the experiments. However, we leave the main discussion about performance of algorithms and lessons learned from the experiments to Section~\ref{sec:discussion}. With as many algorithms and as many graphs as we consider in this paper, it is difficult to present all of the data in a completely satisfying  way. We settled on the following presentation formats. For families of graphs, we plot performance of an algorithm as a time series with an independent variable being the parameter corresponding to the family of graphs and dependent variable being the achieved competitive ratio. The time series allows us to identify regimes of parameters that are easy and that are hard for most algorithms. We list performance of algorithms in those regimes sorted according to their competitive ratios. When we plot the results for these regimes, we also graphically indicate sample standard deviations.  We treat stand-alone and real-world instances differently. We collect performance of all algorithms on all stand-alone instances in one table, and on real-world instances in two tables (one for the random bipartition conversion method and one for the duplicating method). We also use the following notation: we add a letter ``g'' in brackets following an algorithm's name to indicate the greedy version of the algorithm, e.g., \textsc{FeldmanEtAl(g)}. The algorithm's name by itself (e.g., \textsc{FeldmanEtAl}) refers to a non-greedy version of the algorithm. The rest of this section is organized as follows. We describe the results for families of graphs in Subsections~\ref{sec:erdos-renyi-exp}, \ref{sec:left-reg-exp}, \ref{sec:molloy-reed-exp}, and \ref{sec:pref-att-exp}. We present our results for stand-alone graphs in Subsection~\ref{sec:stand-alone-exp} and real-world instances in Subsection~\ref{sec:real-world-exp}. Finally, we finish with a small discussion of running times in Subsection~\ref{sec:runtimes}.

\subsection{Erd\H{o}s-R\'{e}nyi Experiments}
\label{sec:erdos-renyi-exp}

\begin{figure}[H]
\begin{minipage}[b]{0.49\linewidth}
\centering
\includegraphics[width=\textwidth]{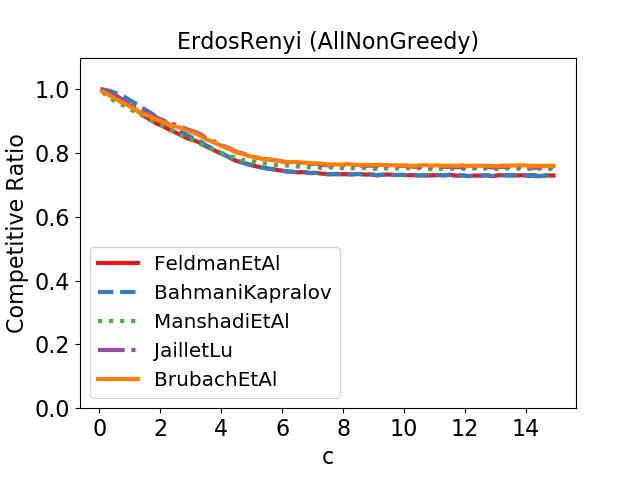}
\caption{Performance of all non-greedy algorithms on Erd\H{o}s-R\'{e}nyi family of graphs.}
\label{fig:non_greedy_erdos_renyi}
\end{minipage}
\begin{minipage}[b]{0.49\linewidth}
\centering
\includegraphics[width=\textwidth]{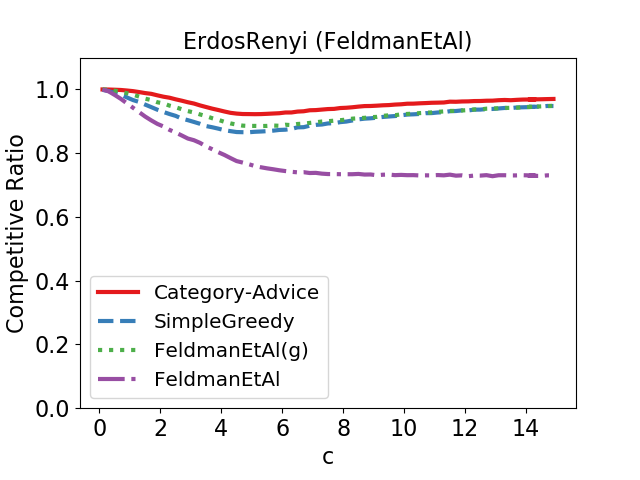}
\caption{Performance of \textsc{FeldmanEtAl} algorithm on Erd\H{o}s-R\'{e}nyi family of graphs.}
\label{fig:feldman_erdos_renyi}
\end{minipage}
\end{figure}

\begin{figure}[H]
\begin{minipage}[b]{0.49\linewidth}
\centering
\includegraphics[width=\textwidth]{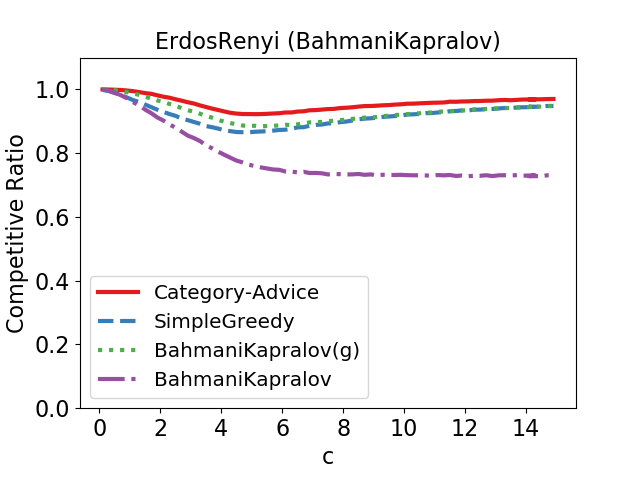}
\caption{Performance of \textsc{BahmaniKapralov} algorithm on Erd\H{o}s-R\'{e}nyi family of graphs.}
\label{fig:bahmani_erdos_renyi}
\end{minipage}
\begin{minipage}[b]{0.49\linewidth}
\centering
\includegraphics[width=\textwidth]{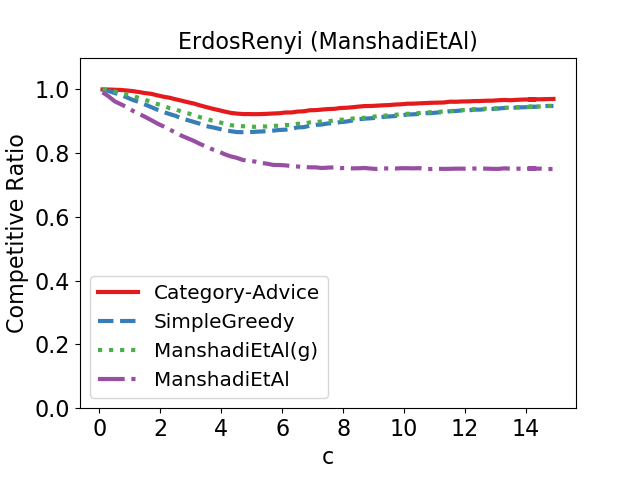}
\caption{Performance of \textsc{ManshadiEtAl} algorithm on Erd\H{o}s-R\'{e}nyi family of graphs.}
\label{fig:manshadi_erdos_renyi}
\end{minipage}
\end{figure}

\begin{figure}[H]
\begin{minipage}[b]{0.49\linewidth}
\centering
\includegraphics[width=\textwidth]{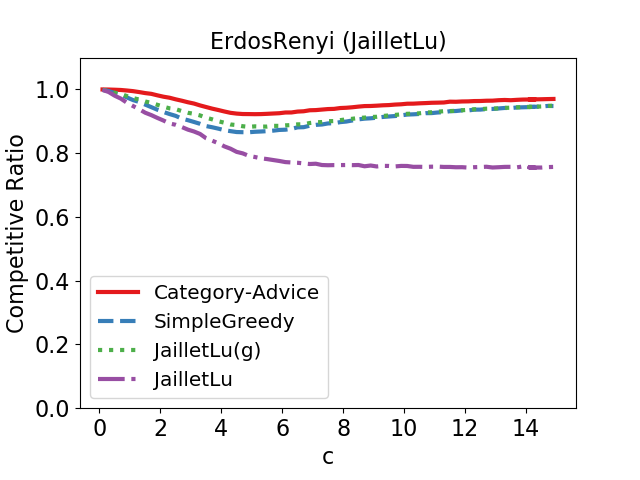}
\caption{Performance of \textsc{JailletLu} algorithm on Erd\H{o}s-R\'{e}nyi family of graphs.}
\label{fig:jaillet_lu_erdos_renyi}
\end{minipage}
\begin{minipage}[b]{0.49\linewidth}
\centering
\includegraphics[width=\textwidth]{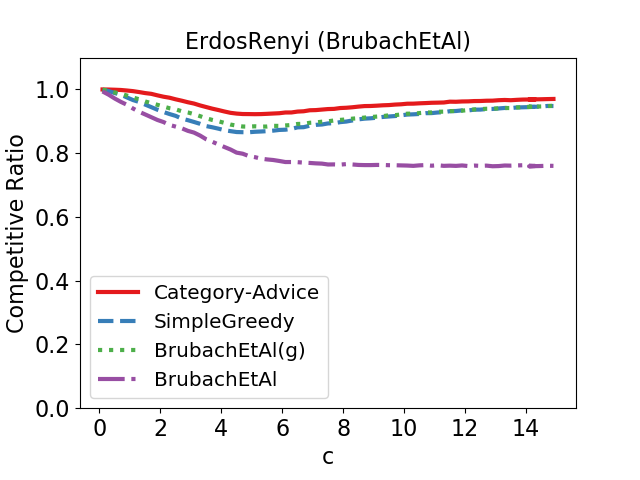}
\caption{Performance of \textsc{BrubachEtAl} algorithm on Erd\H{o}s-R\'{e}nyi family of graphs.}
\label{fig:brubach_erdos_renyi}
\end{minipage}
\end{figure}

\begin{figure}[H]
\begin{minipage}[b]{0.33\linewidth}
\centering
\includegraphics[width=\textwidth]{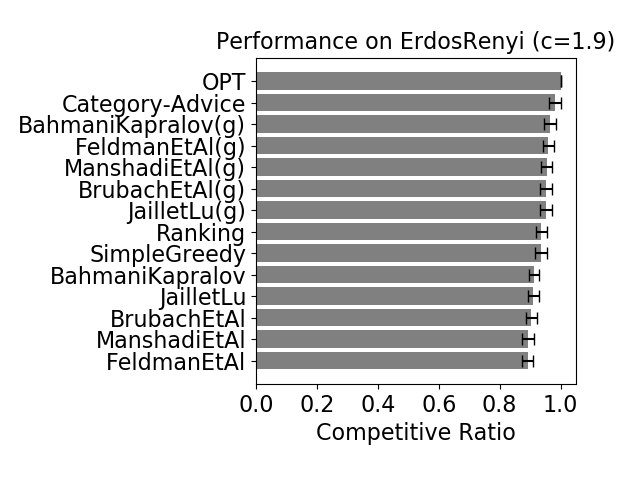}
\caption{Performance of all algorithms on Erd\H{o}s-R\'{e}nyi graph with $c = 1.9$.}
\label{fig:all_erdos_renyi_c_1_9}
\end{minipage}
\begin{minipage}[b]{0.33\linewidth}
\centering
\includegraphics[width=\textwidth]{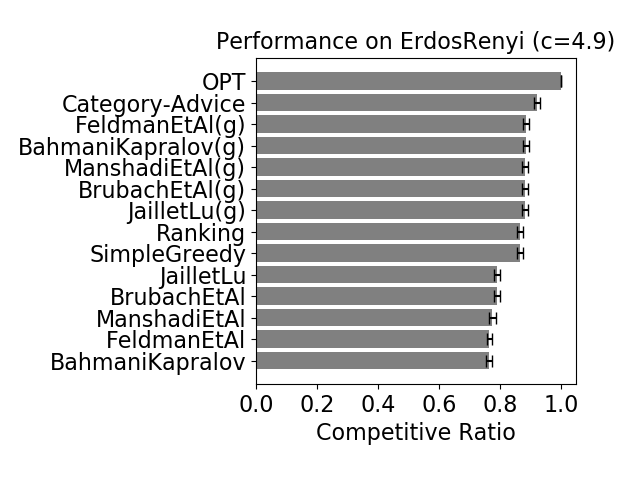}
\caption{Performance of all algorithms on Erd\H{o}s-R\'{e}nyi graph with $c = 4.9$.}
\label{fig:all_erdos_renyi_c_4_9}
\end{minipage}
\begin{minipage}[b]{0.33\linewidth}
\centering
\includegraphics[width=\textwidth]{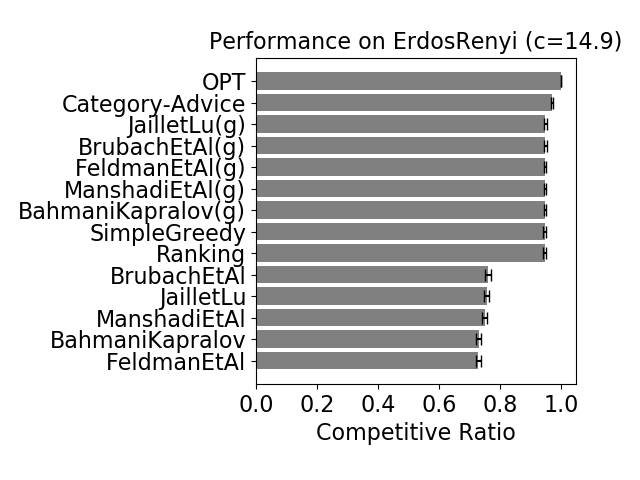}
\caption{Performance of all algorithms on Erd\H{o}s-R\'{e}nyi graph with $c = 14.9$.}
\label{fig:all_erdos_renyi_c_14_9}
\end{minipage}
\end{figure}

The experiments in this section were performed with Erd\H{o}s-R\'{e}nyi type graphs where the number of nodes was fixed to be 1000 on each side, and the parameter $c$ varied from $0.1$ to $14.9$ with a step of $0.2$. For each value of $c$, 100 type graphs were generated. The reported competitive ratios of algorithms are (ratios of) the average values over these 100 trials. In Figure~\ref{fig:non_greedy_erdos_renyi} you can see the time series of performance of all non-greedy algorithms in this experiment. Each non-greedy algorithm is compared with greedy algorithms (including its own counterpart) in Figures~\ref{fig:feldman_erdos_renyi}, \ref{fig:bahmani_erdos_renyi}, \ref{fig:manshadi_erdos_renyi}, \ref{fig:jaillet_lu_erdos_renyi}, and \ref{fig:brubach_erdos_renyi}. We did not plot \textsc{Ranking}, since its behavior in this experiment was analogous to that of \textsc{SimpleGreedy}. Observe that from Figures~\ref{fig:non_greedy_erdos_renyi} and, for example, \ref{fig:feldman_erdos_renyi}, one can infer all other figures. We only show other figures here for completeness, and in the future experiments we shall omit them. Looking at the figures, we observe that there are essentially three regimes of $c$ that are of interest in this experiment: (1) small $c$, i.e., a sparse type graph, regime, (2) ``hard'' values of $c$, where the relative order of algorithms changes, and performance of greedy algorithms experiences a dip, and (3) asymptotic, i.e.,  steady-state, value of $c$, where the performance guarantees of various non-greedy algorithms stabilizes. In order to ``zoom-in'' and see what happens in each of these regimes, we plotted competitive ratios of algorithms in decreasing order (top to bottom) for $c=1.9$ (regime (1)), $c=4.9$ (regime (2)), and $c=14.9$ (regime(3)) in Figures~\ref{fig:all_erdos_renyi_c_1_9}, \ref{fig:all_erdos_renyi_c_4_9}, and \ref{fig:all_erdos_renyi_c_14_9}, respectively.

\subsection{Random Left-Regular Experiments}
\label{sec:left-reg-exp}

\begin{figure}[H]
\begin{minipage}[b]{0.49\linewidth}
\centering
\includegraphics[width=\textwidth]{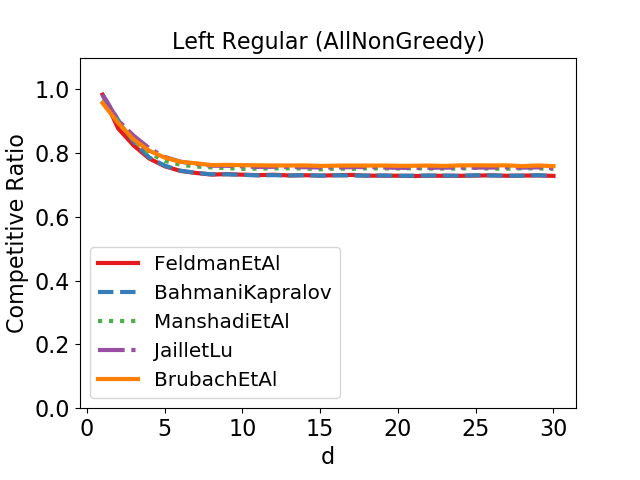}
\caption{Performance of all non-greedy algorithms on Left-Regular family of graphs.}
\label{fig:non_greedy_left_regular}
\end{minipage}
\begin{minipage}[b]{0.49\linewidth}
\centering
\includegraphics[width=\textwidth]{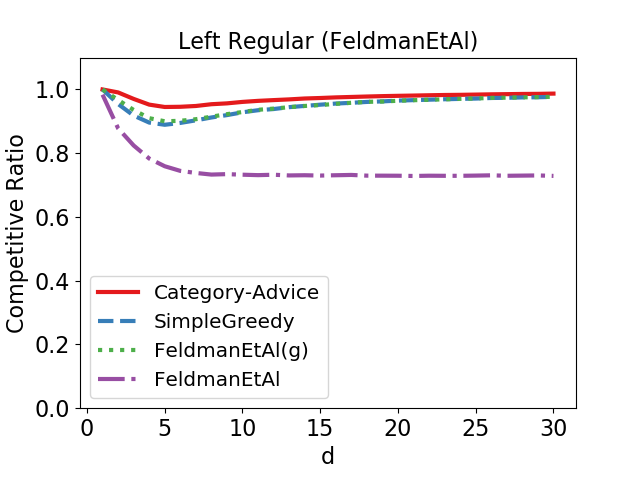}
\caption{Performance of \textsc{FeldmanEtAl} algorithm on Left-Regular family of graphs.}
\label{fig:feldman_left_regular}
\end{minipage}
\end{figure}

\begin{figure}[H]
\begin{minipage}[b]{0.33\linewidth}
\centering
\includegraphics[width=\textwidth]{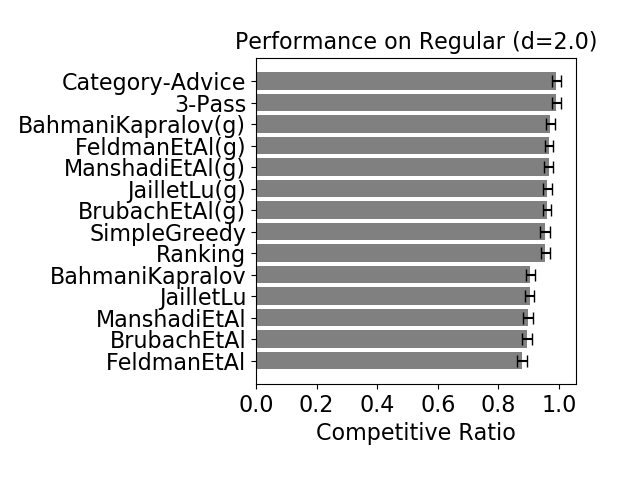}
\caption{Performance of all algorithms on Left-Regular graph with $d = 2$.}
\label{fig:all_left_regular_d_2}
\end{minipage}
\begin{minipage}[b]{0.33\linewidth}
\centering
\includegraphics[width=\textwidth]{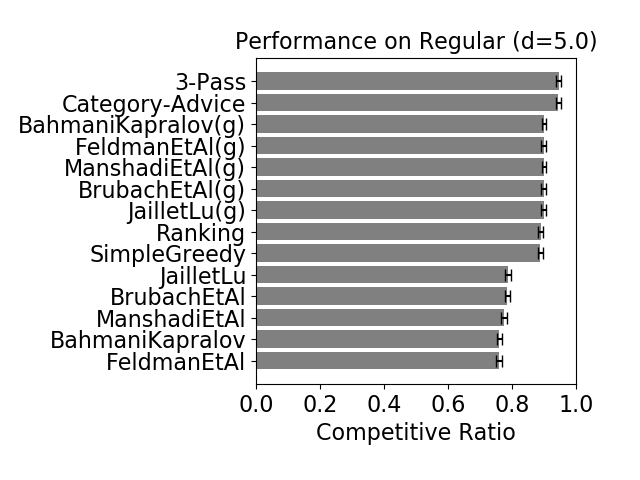}
\caption{Performance of all algorithms on Left-Regular graph with $d=5$.}
\label{fig:all_left_regular_d_5}
\end{minipage}
\begin{minipage}[b]{0.33\linewidth}
\centering
\includegraphics[width=\textwidth]{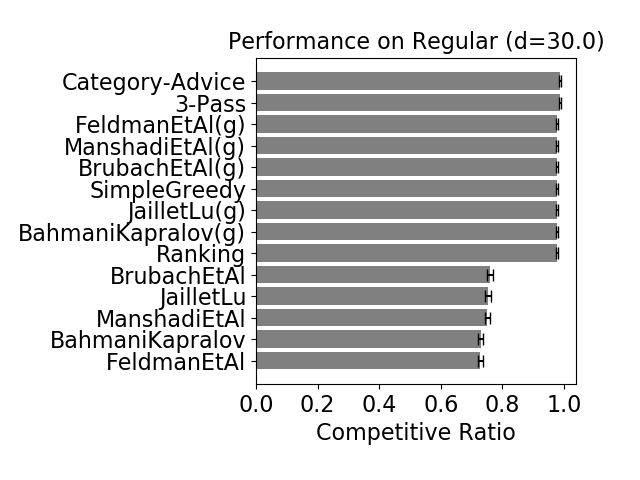}
\caption{Performance of all algorithms on Left-Regular graph with $d=30$.}
\label{fig:all_left_regular_d_30}
\end{minipage}
\end{figure}

The experiments in this subsection are based on type graphs with 1000 nodes on each side, where left-hand-side nodes are of degree $d$ each. As before, results are averaged over $100$ i.i.d. trials. We present time series of all non-greedy algorithms in Figure~\ref{fig:non_greedy_left_regular}, and we present \textsc{FeldmanEtAl} versus greedy algorithms in Figure~\ref{fig:feldman_left_regular}. Figures comparing other non-greedy algorithms with greedy algorithms are omitted, since they are very similar to Figure~\ref{fig:feldman_left_regular}, as discussed at the beginning of Subsection~\ref{sec:erdos-renyi-exp}. We identify three regimes of $d$ that correspond to (1) sparse case ($d=2$), (2) difficult case ($d=5$), and (3) asymptotic case ($d=30$). This is very similar to what we did in Subsection~\ref{sec:erdos-renyi-exp}. The competitive ratios of different algorithms under these regimes are plotted in Figures~\ref{fig:all_left_regular_d_2}, \ref{fig:all_left_regular_d_5}, and \ref{fig:all_left_regular_d_30}.

\subsection{Molloy-Reed Experiments}
\label{sec:molloy-reed-exp}

\begin{figure}[H]
\begin{minipage}[b]{0.49\linewidth}
\centering
\includegraphics[width=\textwidth]{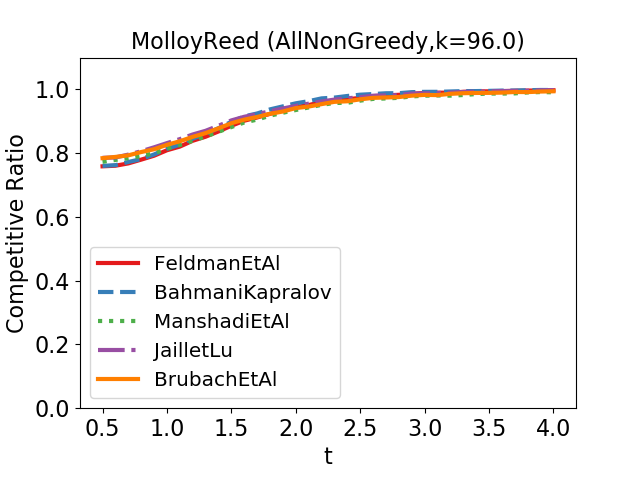}
\caption{Performance (as a function of $\tau$) of all non-greedy algorithms on Molloy-Reed family of graphs with $\kappa=96$.}
\label{fig:non_greedy_molloy_reed_k_96}
\end{minipage}
\begin{minipage}[b]{0.49\linewidth}
\centering
\includegraphics[width=\textwidth]{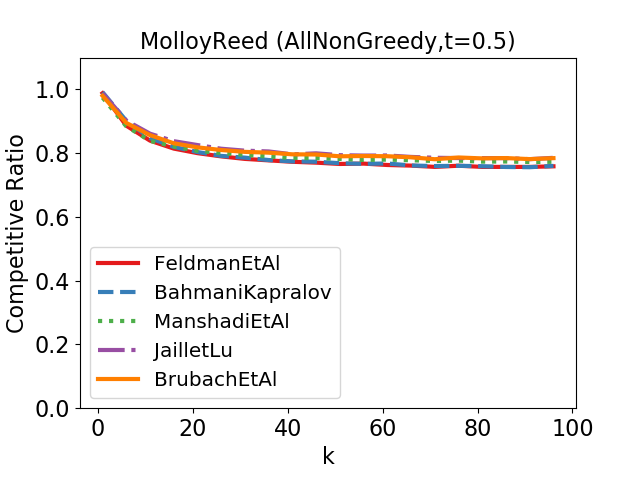}
\caption{Performance (as a function of $\kappa$) of all non-greedy algorithms on Molloy-Reed family of graphs with $\tau=0.5$.}
\label{fig:non_greedy_molloy_reed_t_0_5}
\end{minipage}
\end{figure}

\begin{figure}[H]
\begin{minipage}[b]{0.49\linewidth}
\centering
\includegraphics[width=\textwidth]{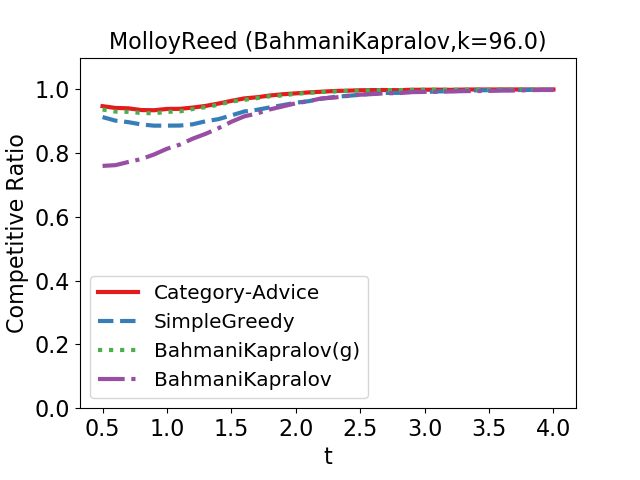}
\caption{Performance (as a function of $\tau$) of \textsc{BahmaniKapralov} algorithm on Molloy-Reed family of graphs with $\kappa=96$.}
\label{fig:bahmani_molloy_reed_k_96}
\end{minipage}
\begin{minipage}[b]{0.49\linewidth}
\centering
\includegraphics[width=\textwidth]{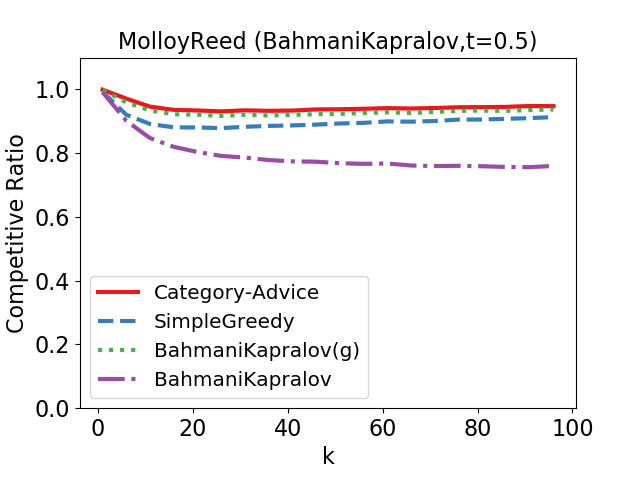}
\caption{Performance (as a function of $\kappa$) of \textsc{BahmaniKapralov} algorithm on Molloy-Reed family of graphs with $\tau=0.5$.}
\label{fig:bahmani_molloy_reed_t_0_5}
\end{minipage}
\end{figure}

\begin{figure}[H]
\begin{minipage}[b]{0.49\linewidth}
\centering
\includegraphics[width=\textwidth]{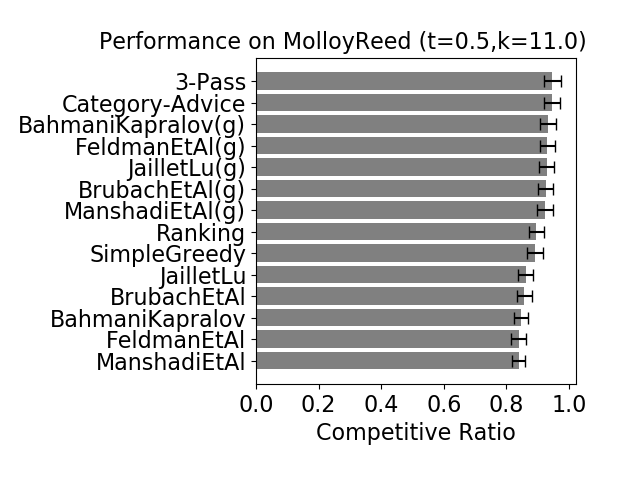}
\caption{Performance of all algorithms on Molloy-Reed graph with $\tau=0.5,\kappa=11$.}
\label{fig:all_molloy_reed_t_0_5_k_11}
\end{minipage}
\begin{minipage}[b]{0.49\linewidth}
\centering
\includegraphics[width=\textwidth]{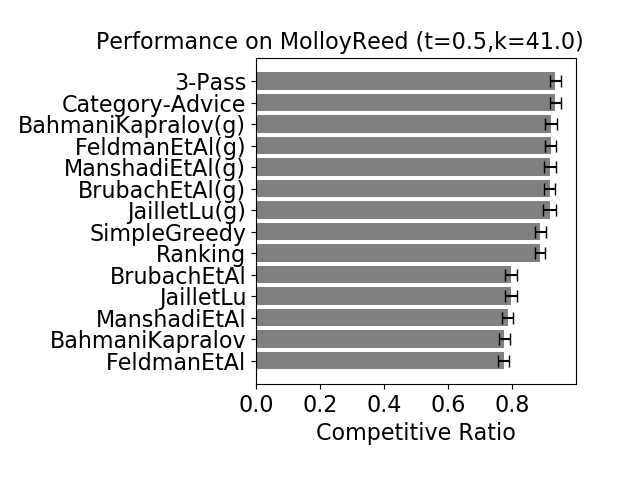}
\caption{Performance of all algorithms on Molloy-Reed graph with $\tau=0.5,\kappa=41$.}
\label{fig:all_molloy_reed_t_0_5_k_41}
\end{minipage}
\end{figure}

\begin{figure}[H]
\begin{minipage}[b]{0.49\linewidth}
\centering
\includegraphics[width=\textwidth]{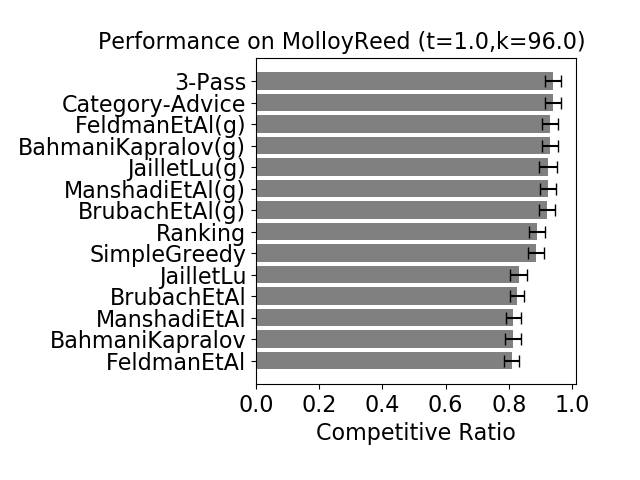}
\caption{Performance of all algorithms on Molloy-Reed graph with $\tau=1.0,\kappa=96$.}
\label{fig:all_molloy_reed_t_1_0_k_96}
\end{minipage}
\begin{minipage}[b]{0.49\linewidth}
\centering
\includegraphics[width=\textwidth]{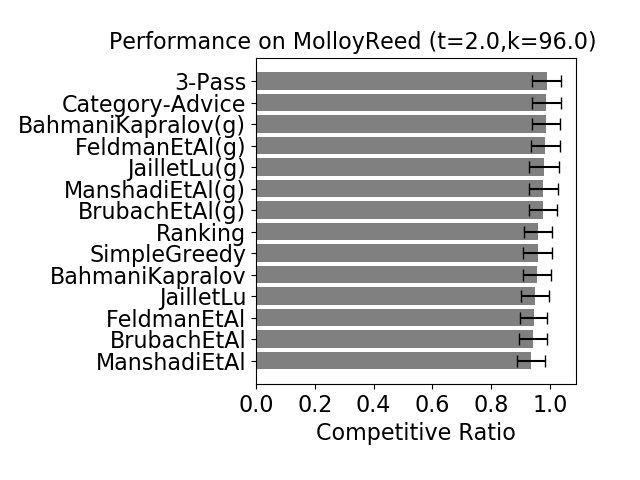}
\caption{Performance of all algorithms on Molloy-Reed graph with $\tau=2.0,\kappa=96$.}
\label{fig:all_molloy_reed_t_2_0_k_96}
\end{minipage}
\end{figure}

Molloy-Reed family of graphs has two parameters: $\tau$ and $\kappa$. Thus, we generated a whole grid of results. More specifically, for each value of $\tau$ from $0.5$ to $4.0$ with a step of $0.1$ and for each value of $\kappa$ from $1$ to $96$ with a step of $5$ we generated $100$ Molloy-Reed graphs with those values of $\tau$ and $\kappa$ and averaged competitive ratios of algorithms over these $100$ runs. Since plotting 3-dimensional time series is awkward, we present $\tau-$ and $\kappa-$slices of the resulting grid for values of $\tau$ and $\kappa$ that exhibit more interesting behavior. We show how competitive ratios of non-greedy algorithms look like as a function of $\tau$ when $\kappa$ is fixed to $96$ in Figure~\ref{fig:non_greedy_molloy_reed_k_96}, and as a function of $\kappa$ when $\tau$ is fixed to $0.5$ in Figure~\ref{fig:non_greedy_molloy_reed_t_0_5}. Time series comparing the non-greedy version of \textsc{BahmaniKapralov} with greedy algorithms for the respective scenarios are shown in Figures~\ref{fig:bahmani_molloy_reed_k_96} and \ref{fig:bahmani_molloy_reed_t_0_5}. As in other subsections, comparisons of other non-greedy algorithms with greedy algorithms look very similar, so we omit them. For $\tau=0.5$ we identify two regimes: difficult regime for greedy algorithms, where $\kappa=11$; and a steady-state regime, where $\kappa=41$. We ``zoom in'' to show competitive ratios of algorithms for these two regimes in Figures~\ref{fig:all_molloy_reed_t_0_5_k_11} and \ref{fig:all_molloy_reed_t_0_5_k_41}. Similarly, the two regimes for $\kappa=96$ are when $\tau=1.0$ and when $\tau=2.0$. Those are depicted in Figures~\ref{fig:all_molloy_reed_t_1_0_k_96} and \ref{fig:all_molloy_reed_t_2_0_k_96}.

\subsection{Preferential Attachment Bigraph Experiments}
\label{sec:pref-att-exp}

\begin{figure}[H]
\begin{minipage}[b]{0.49\linewidth}
\centering
\includegraphics[width=\textwidth]{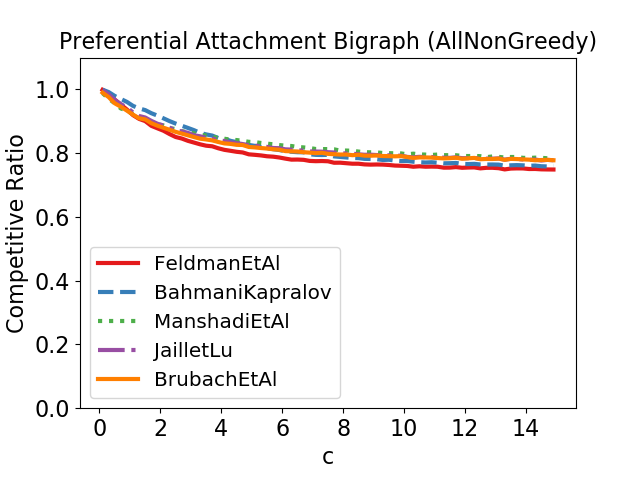}
\caption{Performance of all non-greedy algorithms on Preferential Attachment Bigraph family of graphs.}
\label{fig:non_greedy_pref_att}
\end{minipage}
\begin{minipage}[b]{0.49\linewidth}
\centering
\includegraphics[width=\textwidth]{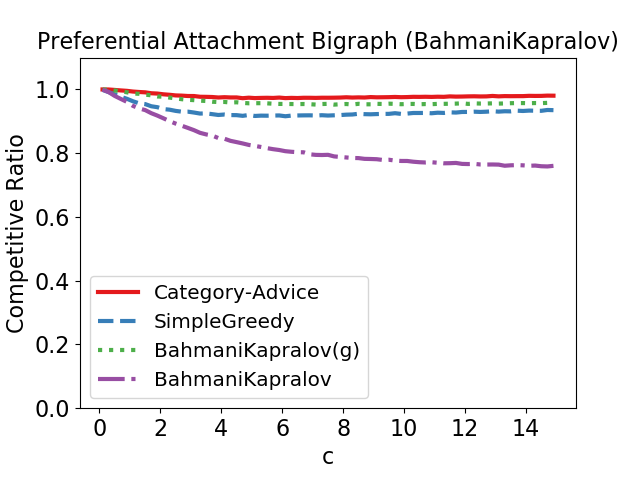}
\caption{Performance of \textsc{BahmaniKapralov} algorithm on Preferential Attachment Bigraph family of graphs.}
\label{fig:bahmani_pref_att}
\end{minipage}
\end{figure}

\begin{figure}[H]
\begin{minipage}[b]{0.33\linewidth}
\centering
\includegraphics[width=\textwidth]{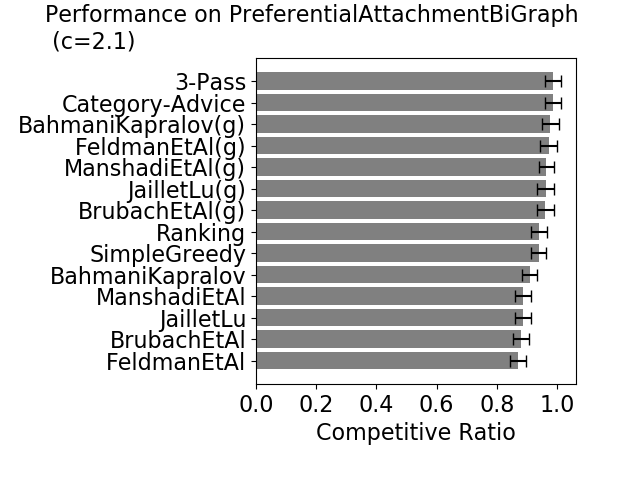}
\caption{Performance of all algorithms on Preferential Attachment Bigraph  graph with $c = 2.1$.}
\label{fig:all_pref_att_c_2_1}
\end{minipage}
\begin{minipage}[b]{0.33\linewidth}
\centering
\includegraphics[width=\textwidth]{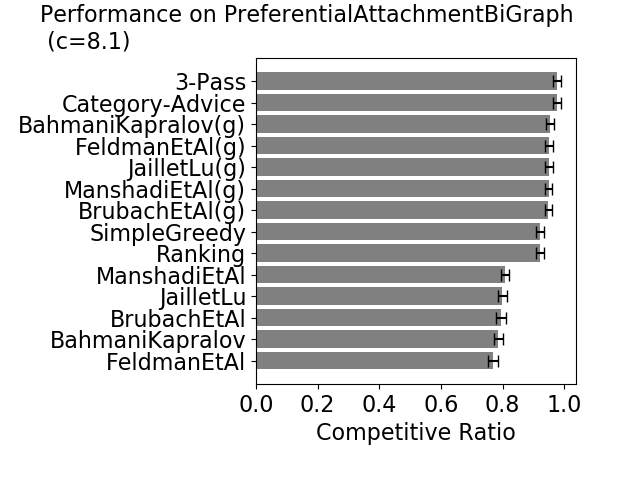}
\caption{Performance of all algorithms on Preferential Attachment Bigraph  graph with $c = 8.1$.}
\label{fig:all_pref_att_c_8_1}
\end{minipage}
\begin{minipage}[b]{0.33\linewidth}
\centering
\includegraphics[width=\textwidth]{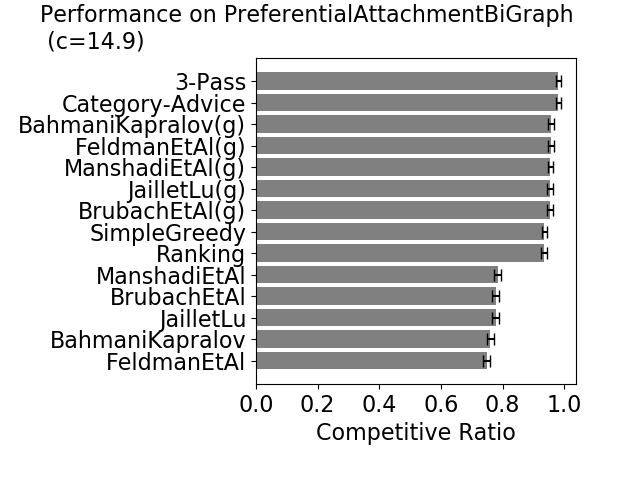}
\caption{Performance of all algorithms on Preferential Attachment Bigraph  graph with $c = 14.9$.}
\label{fig:all_pref_att_c_14_9}
\end{minipage}
\end{figure}

The experiments in this subsection are based on type graphs with 1000 nodes on each side, where types are generated via the preferential attachment method with a single parameter $c$. The values of $c$ range from $0.1$ to $14.9$ with a step of $0.1$. All results are averaged over $100$ i.i.d. trials. We present time series of all non-greedy algorithms in Figure~\ref{fig:non_greedy_pref_att}, and we present \textsc{BahmaniKapralov} versus greedy algorithms in Figure~\ref{fig:bahmani_pref_att}. Figures comparing other non-greedy algorithms with greedy algorithms are omitted, since they can be inferred from the given two figures, as discussed at the beginning of Subsection~\ref{sec:erdos-renyi-exp}. We identify three regimes of $c$ that correspond to (1) sparse case ($c=2.1$), (2) intermediate case ($c=8.1$), and (3) asymptotic case ($c=14.9$); this is similar to Subsection~\ref{sec:erdos-renyi-exp}. The competitive ratios of different algorithms are plotted in Figures~\ref{fig:all_pref_att_c_2_1}, \ref{fig:all_pref_att_c_8_1}, and \ref{fig:all_pref_att_c_14_9}.

\subsection{Stand-Alone Graphs}
\label{sec:stand-alone-exp}

In this subsection, we collect competitive ratios of all algorithms considered in this paper on all stand-alone graphs, as discussed in Section~\ref{sec:experimental_setup}. Since some of the algorithms and graph constructions are randomized, we show results that are averaged over $100$ trials. See Table~\ref{tab:all_stand_alone} for the summary.

\begin{table}[H]
\centering
\begin{tabular}{ |c||c|c|c|c|c|c|c|c|}
\hline
Algorithm &UT &MH &FH &FewG &ManyG &Rope &Hexa &Zipf\\
\hline
\hline
FeldmanEtAl & 0.76 & 0.76 & 0.67 & 0.77 & 0.80 & 0.92 & 0.75 & 0.86\\
\hline
FeldmanEtAl(g) & 0.90 & 0.87 & 0.87 & 0.89 & 0.92 & 0.99 & 0.89 & 0.96\\
\hline
BahmaniKapralov & 0.76 & 0.76 & 0.80 & 0.77 & 0.79 & 0.92 & 0.75 & 0.93\\
\hline
BahmaniKapralov(g) & 0.90 & 0.87 & 0.93 & 0.89 & 0.92 & 0.99 & 0.89 & 0.98\\
\hline
ManshadiEtAl & 0.77 & 0.79 & 0.84 & 0.78 & 0.80 & 0.92 & 0.77 & 0.90\\
\hline
ManshadiEtAl(g) & 0.89 & 0.87 & 0.96 & 0.89 & 0.92 & 0.99 & 0.89 & 0.96\\
\hline
JailletLu & 0.78 & 0.80 & 0.78 & 0.80 & 0.82 & 0.93 & 0.78 & 0.87\\
\hline
JailletLu(g) & 0.90 & 0.87 & 0.87 & 0.89 & 0.92 & 0.99 & 0.89 & 0.94\\
\hline
BrubachEtAl & 0.78 & 0.81 & 0.78 & 0.80 & 0.82 & 0.94 & 0.78 & 0.87\\
\hline
BrubachEtAl(g) & 0.91 & 0.87 & 0.92 & 0.89 & 0.92 & 0.99 & 0.89 & 0.95\\
\hline
SimpleGreedy & 0.66 & 0.87 & 0.91 & 0.86 & 0.90 & 0.99 & 0.86 & 0.87\\
\hline
Ranking & 0.92 & 0.88 & 0.94 & 0.87 & 0.91 & 0.99 & 0.87 & 0.93\\
\hline
Category-Advice & 0.76 & 0.95 & 0.99 & 0.92 & 0.95 & 1.00 & 0.92 & 0.97\\
\hline
3-Pass & 0.77 & 0.95 & 0.99 & 0.92 & 0.95 & 1.00 & 0.92 & 0.97\\
\hline
OPT & 1.00 & 1.00 & 1.00 & 1.00 & 1.00 & 1.00 & 1.00 & 1.00\\
\hline
\end{tabular}
\caption{Performance of algorithms on stand-alone graphs.}
    \label{tab:all_stand_alone}
\end{table}

\subsection{Real-World Instances}
\label{sec:real-world-exp}

In this subsection, we collect competitive ratios of our algorithms on graphs based on  real-world applications, as discussed in Section~\ref{sec:experimental_setup}. That is, since these  application graphs are not bipartite, we consider two methods of converting them into bipartite graph: the random bipartition method, and the duplicating method. Since some of the algorithms and the random bipartition conversion method  are randomized, we show results that are averaged over $100$ trials. See Table~\ref{tab:real_world_inst_rb} for the summary of results for the random bipartition method, and Table~\ref{tab:real_world_inst_dup} for the duplicating method.

\begin{table}[H]
\begin{center}
\begin{tabular}{ |c||c|c|c|c|c|c|}
\hline
Algorithm &Caltech36 &Reed98 &CE-GN &CE-PG &beause &mbeaflw\\
\hline
\hline
FeldmanEtAl & 0.78 & 0.78 & 0.78 & 0.81 & 0.76 & 0.74\\
\hline
FeldmanEtAl(g) & 0.91 & 0.91 & 0.94 & 0.96 & 0.94 & 0.95\\
\hline
BahmaniKapralov & 0.80 & 0.81 & 0.84 & 0.89 & 0.80 & 0.76\\
\hline
BahmaniKapralov(g) & 0.92 & 0.92 & 0.96 & 0.98 & 0.96 & 0.96\\
\hline
ManshadiEtAl & 0.81 & 0.81 & 0.84 & 0.87 & 0.81 & 0.79\\
\hline
ManshadiEtAl(g) & 0.91 & 0.91 & 0.96 & 0.97 & 0.96 & 0.96\\
\hline
JailletLu & 0.81 & 0.81 & 0.80 & 0.82 & 0.78 & 0.77\\
\hline
JailletLu(g) & 0.91 & 0.91 & 0.94 & 0.96 & 0.95 & 0.96\\
\hline
BrubachEtAl & 0.81 & 0.81 & 0.81 & 0.83 & 0.79 & 0.77\\
\hline
BrubachEtAl(g) & 0.91 & 0.91 & 0.94 & 0.95 & 0.94 & 0.95\\
\hline
SimpleGreedy & 0.87 & 0.87 & 0.93 & 0.94 & 0.94 & 0.95\\
\hline
Ranking & 0.86 & 0.87 & 0.93 & 0.94 & 0.94 & 0.95\\
\hline
Category-Advice & 0.92 & 0.93 & 0.97 & 0.98 & 0.97 & 0.97\\
\hline
3-Pass & 0.92 & 0.93 & 0.97 & 0.98 & 0.97 & 0.97\\
\hline
OPT & 1.00 & 1.00 & 1.00 & 1.00 & 1.00 & 1.00\\
\hline
\end{tabular}
\end{center}
  \caption{Performance of algorithms on real-life instances transformed into bipartite instances via the random-bipartition method.}
    \label{tab:real_world_inst_rb}
\end{table}

\begin{table}[H]
\begin{center}
\begin{tabular}{ |c||c|c|c|c|c|c|}
\hline
Algorithm &Caltech36 &Reed98 &CE-GN &CE-PG &beause &mbeaflw\\
\hline
\hline
FeldmanEtAl & 0.77 & 0.77 & 0.77 & 0.80 & 0.74 & 0.73\\
\hline
FeldmanEtAl(g) & 0.90 & 0.90 & 0.95 & 0.95 & 0.94 & 0.97\\
\hline
BahmaniKapralov & 0.78 & 0.78 & 0.82 & 0.88 & 0.76 & 0.75\\
\hline
BahmaniKapralov(g) & 0.91 & 0.91 & 0.97 & 0.98 & 0.95 & 0.97\\
\hline
ManshadiEtAl & 0.79 & 0.78 & 0.84 & 0.86 & 0.78 & 0.77\\
\hline
ManshadiEtAl(g) & 0.90 & 0.90 & 0.96 & 0.97 & 0.95 & 0.96\\
\hline
JailletLu & 0.79 & 0.79 & 0.79 & 0.81 & 0.77 & 0.76\\
\hline
JailletLu(g) & 0.90 & 0.90 & 0.95 & 0.96 & 0.95 & 0.97\\
\hline
BrubachEtAl & 0.80 & 0.79 & 0.80 & 0.82 & 0.77 & 0.76\\
\hline
BrubachEtAl(g) & 0.91 & 0.91 & 0.94 & 0.95 & 0.95 & 0.97\\
\hline
SimpleGreedy & 0.72 & 0.72 & 0.95 & 0.95 & 0.91 & 0.94\\
\hline
Ranking & 0.86 & 0.86 & 0.93 & 0.94 & 0.94 & 0.97\\
\hline
Category-Advice & 0.82 & 0.83 & 0.98 & 0.99 & 0.96 & 0.97\\
\hline
3-Pass & 0.83 & 0.84 & 0.98 & 0.99 & 0.96 & 0.97\\
\hline
OPT & 1.00 & 1.00 & 1.00 & 1.00 & 1.00 & 1.00\\
\hline
\end{tabular}
\end{center}
  \caption{Performance of algorithms on real-life instances transformed into bipartite instances via the duplicating method.}
    \label{tab:real_world_inst_dup}
\end{table}

\subsection{Running Times}
\label{sec:runtimes}

\begin{figure}[H]
\centering
\includegraphics[width=43em]{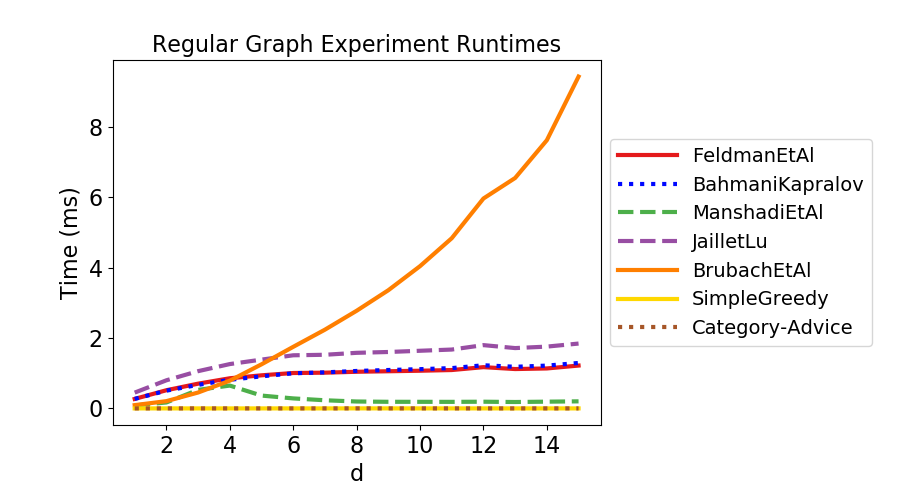}
\caption{Running times left-side regular family of graphs.}
\label{fig:regular_times}
\end{figure}

In this section we describe our experimental findings about running times of the algorithms under consideration. The goal is to see how running times scale with the number of edges. For that purpose we present running times from  random left- regular graph experiments. The edge density for the other data set  families in Section~\ref{subsec:families} are also controlled by a single parameter and the run times for experiments for these data sets behave exactly the same way. Thus, the conclusions we derive for the random left regular graph experiments are quite general and are expected to carry over to other scenarios. In the random left regular graph experiment, the number of online nodes is fixed to be 1000, same as the number of offline nodes. The neighborhood of each online node is decided by selecting a subset of neighbors of size $d$ uniformly at random. Thus, as $d$ increases, the total number of edges increases. Figure~\ref{fig:regular_times} shows how the running times of the  various algorithms scale with $d$. 

For ease of presentation, we omit greedy versions of complicated algorithms (i.e., \textsc{FeldmanEtAl}, \textsc{BahmaniKapralov}, \textsc{ManshadiEtAl}, \textsc{JailletLu}, \textsc{BrubachEtAl}). In our experiments, greedy versions had the same runtimes as their non-greedy versions, because turning an algorithm into a greedy one has virtually no overhead. Runtimes of complicated algorithms are dominated by their preprocessing stages. Therefore, the more complicated is the preprocessing stage, the slower is the algorithm. Greedy-like algorithms have either no or minimal preprocessing, e.g. \textsc{SimpleGreedy} and \textsc{Category-Advice}, and thus are the fastest algorithms, as expected. \textsc{BahmaniKapralov} is essentially \textsc{FeldmanEtAl} with some additional preprocessing steps, thus the two algorithms behave similarly with \textsc{BahmaniKapralov} being slightly slower. \textsc{JailletLu} is slower still, but not by much. The behavior of \textsc{ManshadiEtAl} might seem mysterious at first as  the running time increases sharply until $d=4$ and then suddenly drops almost matching the greedy runtime. To explain this behavior, we recall how the preprocessing step of \textsc{ManshadiEtAl} works. It samples 100 graphs from the distribution specified by the type graph and solves each of the samples optimally. Thus, the runtime depends not only on the density of the edges, but also on how easy it is to solve a sample optimally. The runtime plot suggests that the case $d=4$ is the hardest to solve optimally (among integral $d$). Clearly, as $d$ increases, it becomes easier and easier to find a perfect matching in the samples. This leads to a faster runtime when $d > 4$. Of course, we applied a simplification where we fixed the number of samples used by \textsc{ManshadiEtAl} in the preprocessing stage to be 100. In practice, one would have to adjust the number of samples with the density of the graph with denser type graphs requiring far more samples. Thus, one would expect the run time of \textsc{ManshadiEtAl} to scale much worse than what is suggested by our figure. So far, our plot suggests that the runtimes of simple greedy-like algorithms scale linearly with the number of edges, while the runtimes of other more complicated algorithms scale like small polynomials. When it comes to the last complicated algorithm, \textsc{BrubachEtAl}, the runtime scales exponentially with $d$. The runtime of \textsc{BrubachEtAl} is dominated by the part of the preprocessing stage that corresponds to solving the LP. The number of constraints in the  LP of \textsc{BrubachEtAl} is asymptotically larger than the number of edges (assuming $|V| = o(|E|)$). More specifically, let $e(r)$ denote the number of edges incident on a node $r \in R$ in the given type graph. The number of constraints in the LP of \textsc{BrubachEtAl} is at least $\sum_{r \in R} e(r)^2 \ge |E|^2/n$ (by Cauchy-Schwarz). Using the simplex method from the  GNU Linear Programming Toolkit even with moderately dense graphs (e.g., $d = 100$) already results in excessive runtimes and in memory consumption of over 6GB.  One could potentially try to optimize this step, possibly applying interior-point methods to very large instances and designing new heuristics to speed up this computation. We suspect that such efforts would not be worth it; it does not seem feasible to run \textsc{BrubachEtAl} on very large instances (e.g., number of edges on the order of tens of millions) and furthermore simpler methods either match or surpass its performance in terms of the competitive ratio on many instances that we consider in this study.

\section{Discussion}
\label{sec:discussion}
As seen in  Figure~\ref{fig:non_greedy_erdos_renyi}, in the \textbf{Erd\H{o}s-R\'{e}nyi} family of graphs, all algorithms exhibit somewhat similar performance. All algorithms perform much better than their theoretical worst-case guarantees.  This is expected because of the randomness in the input. Figures~\ref{fig:feldman_erdos_renyi}-\ref{fig:brubach_erdos_renyi} showcase the experimental competitive ratios of each algorithm along with its greedy version, \textsc{SimpleGreedy}, and \textsc{Category-Advice} (2-Pass). All algorithms seem to follow similar trends. More specifically, the non-greedy versions show a drop in performance as $c$ increases. This is to be expected since non-greedy algorithms ignore a certain fraction of the input while the offline optimum increases when the graph is getting denser. Greedy algorithms always perform close to optimum when the graph is very sparse or very dense and this behavior is evident as the greedy versions achieve a global minimum around $c=4.9$. For a theoretical explanation of this behavior see \cite{MastinJ2013} and \cite{BorodinKP2018}. Figures \ref{fig:all_erdos_renyi_c_1_9}-\ref{fig:all_erdos_renyi_c_14_9} show an experimental ranking of the algorithms before, around, and after the global minimum respectively. What stands out the most is that even the simplest greedy algorithms (\textsc{SimpleGreedy} and \textsc{Ranking}) always outperform the more sophisticated non-greedy algorithms that make use of the type graph, while the greedy versions of the latter perform only slightly better than simple greedy ones. Interestingly, \textsc{Category-Advice} is always the best performing algorithm. For $c=14.9$ the experimental ranking of the non-greedy algorithms is consistent with table \ref{tab:algos_ranking}. For $c=1.9$ and $c=4.9$ \textsc{BahmaniKapralov} is the best and worst non-greedy algorithm respectively, while in both cases \textsc{JailletLu} is doing slightly better than \textsc{BrubachEtAl}. It's worth noting that the proven competitive ratios of \textsc{JailletLu} and \textsc{BrubachEtAl} only differ by 0.0006 and the number of simulations required to achieve a low enough error margin is beyond our computational resources. We also note that \textsc{3-Pass} is always guaranteed to perform at least as well as \textsc{Category-Advice}, but sometimes our plots list \textsc{Category-Advice}  above \textsc{3-Pass}. In these instances, the two algorithms gave identical performance, and the plot generating procedure broke ties in favour of \textsc{Category-Advice}. In all our experiments \textsc{3-Pass} was  either identical to \textsc{Category-Advice} or gave minuscule improvements. This is explained by an already impressive performance of \textsc{Category-Advice}. When the instance graph has a near-perfect matching, one can expect the first and second pass to cover almost all offline nodes. The third pass differs from the first two only in its behavior on the nodes that were not matched in either of the first two passes. If the first two passes already cover all or almost all nodes, then the third pass doesn't actually do anything, and this is what we observed.

The \textbf{Random Regular} family of graphs paints a similar picture. The results are almost identical for \textit{Left Regular} and \textit{Right Regular}, so we only present the former. The performance of all non-greedy algorithms is depicted in Figure \ref{fig:non_greedy_left_regular}. The main difference compared to the Erd\H{o}s-R\'{e}nyi family is that the algorithms converge a bit faster as the graph gets denser. The results of non-greedy \textsc{Feldman} algorithm compared to its greedy version and other greedy algorithms are shown in Figure~\ref{fig:feldman_left_regular}. Other non-greedy algorithms exhibit similar behavior. We see that the greedy algorithms achieve a global minimum around $d=5$ and that there seems to be a slightly bigger gap between the non-greedy algorithms and their greedy versions than in the Erd\H{o}s-R\'{e}nyi experiment. As seen in Figures \ref{fig:all_left_regular_d_2}-\ref{fig:all_left_regular_d_30}, for $d=5$ and $d=30$, the ranking of the non-greedy algorithms agrees with their respective theoretical guarantees, while for $d=2$ \textsc{BahmaniKapralov} performs the best, and \textsc{BrubachEtAl} is second worst after \textsc{FeldmanEtAl}. \textsc{SimpleGreedy} and \textsc{Ranking} are consistently better than all non-greedy algorithms, while greedy algorithms that use information from the type graph perform marginally better than \textsc{SimpleGreedy} and \textsc{Ranking} in all cases except for $d=30$ where \textsc{SimpleGreedy} beats \textsc{JailletLu(g)} and \textsc{BrubachEtAl(g)}.

For the \textbf{Molloy-Reed} family of graphs, Figures \ref{fig:non_greedy_molloy_reed_k_96}, \ref{fig:non_greedy_molloy_reed_t_0_5} show the performance of non-greedy algorithms as functions of $\tau$ and $\kappa$, respectively. The behavior of all algorithms is again very similar. We can see that as $\kappa$ increases, the graph is getting denser, which results in an expected drop in performance as more nodes that could be included in an optimal matching are being rejected. On the contrary, as $\tau$ increases, the graph is getting less dense and the performance increases. In fact, as shown in the indicative plot of \textsc{BahmaniKapralov} (Figure \ref{fig:bahmani_molloy_reed_k_96}), as $\tau$ increases and the graph becomes sparse, non-greedy algorithms achieve performance close to that of greedy ones. The experimental performance-based ranking of algorithms in various settings is shown in Figures \ref{fig:all_molloy_reed_t_0_5_k_11}-\ref{fig:all_molloy_reed_t_2_0_k_96}. \textsc{Ranking} and \textsc{SimpleGreedy} consistently outperform non-greedy algorithms but the greedy versions of algorithms that use the type graph achieve slightly better results. The ranking of non-greedy algorithms varies a lot among different parameter settings. As opposed to other graphs, \textsc{FeldmanEtAl} does not always come last, and it even outperforms \textsc{BrubachEtAl} in the setting of $(\tau = 2,\kappa = 96.0)$.

Results for the \textbf{Preferential Attachment Bigraph} family (Figures \ref{fig:non_greedy_pref_att}-\ref{fig:all_pref_att_c_14_9}) agree with the patterns observed thus far. The greedy version of only one of the non-greedy algorithms is presented in a plot (Figure \ref{fig:bahmani_pref_att}), and the rest are similar.
\textsc{SimpleGreedy} and \textsc{Ranking} do better than all non-greedy algorithms but do not perform as well as the greedy versions of algorithms using the type graph. Interestingly, \textsc{BahmaniKapralov(g)} followed by \textsc{FeldmanEtAl(g)} are the two best performing online algorithms. For $c=8.1$ and $c=14.9$, \textsc{ManshadiEtAl} is the best non-greedy algorithm, whereas for $c=2.1$ it comes second to \textsc{BahmaniKapralov}.

Similar trends appear in the results for stand-alone graphs (table \ref{tab:all_stand_alone}), with algorithms performing much better than their worst case guarantees. One exception is graph \textbf{UT}, where \textsc{SimpleGreedy} is the worst performing algorithm. This is to be expected as that is the worst case graph for that algorithm. Besides \textsc{SimpleGreedy}, \textsc{Category-Advice} and \textsc{3-Pass} algorithms also achieve low performance on graph UT. On the other hand, \textsc{Ranking} is the best algorithm for UT, even though in the adversarial setting that is its worst-case graph. Moreover, it seems that 0.78 is an upper bound of all other non-greedy algorithms on UT, indicating that this graph might be a useful theoretical benchmark.
Graph \textbf{MH} is a hard instance that produces the best known upper bound for the known i.i.d. \emph{input model} with integral types. Algorithm \textsc{Ranking} achieves the best online performance, while \textsc{Category-Advice} and \textsc{3-Pass} result in a substantial improvement. The exceptionally low performance of \textsc{FeldmanEtAl} on graph \textbf{FH} verifies it as its worst-case graph. What is also interesting about FH is that the experimental performance-based ranking of the algorithms is not consistent with the ranking of table \ref{tab:algos_ranking}. Specifically, the best non-greedy algorithm is \textsc{ManshadiEtAl} with 0.84, followed by \textsc{BahmaniKapralov} and the remaining algorithms in their usual ordering. \textbf{Rope} seems to be the easiest class that all algorithms can handle quite well. The worst performance on graph Rope is 0.92 achieved by \textsc{FeldmanEtAl}, \textsc{BahmaniKapralov} and \textsc{ManshadiEtAl}. The second easiest graph is \textbf{Zipf} where the worst performing algorithm is \textsc{FeldmanEtAl} with a competitive ratio of 0.86.
On graphs \textbf{FewG} and \textbf{ManyG}, the ranking of the non-greedy algorithms algorithms based on their performance follows their ranking based on their worst-case analysis with a small exception of \textsc{FeldmanEtAl} beating \textsc{BahmaniKapralov} on \textbf{ManyG} by 0.01. Overall, \textbf{ManyG} appears to be an easy graph, which might be due to a quite uniform distribution of edges over the offline nodes. \textbf{Hexa} appears to be an instance of similar hardness to \textbf{MH}. The best performing algorithms get 0.89, only slightly better than the best performance on \textbf{MH} (0.88), while the worst algorithm (\textsc{FeldmanEtAl} with 0.75) is slightly worse than the worst performance on \textbf{MH} (0.76). It is also worth noting that the multiple-pass offline algorithms do not result in a performance increase as big as on graph \textbf{MH}.

Overall, the ranking of the non-greedy algorithms as presented in Table~\ref{tab:algos_ranking} remains fairly consistent in the stand-alone graphs, except for \textbf{FH} and \textbf{Zipf} where \textsc{ManshadiEtAl} and \textsc{BahmaniKapralov} take the lead. Additionally, algorithms \textsc{Ranking} and \textsc{SimpleGreedy} always outperform the non-greedy algorithms that make use of the type graph, with the former always being slightly better, and with the only exception being graph \textbf{UT}, where \textsc{SimpleGreedy} experiences a drop in performance. We also find that the greedy versions of algorithms that use type graph information are in some cases better than \textsc{Ranking} (\textbf{FewG}, \textbf{ManyG}, \textbf{Hexa}, \textbf{Zipf}), but that is not always the case. In just two graphs, namely \textbf{UT} and \textbf{Zipf}, there exist online algorithms that beat the offline multiple-pass algorithms.

As shown in Tables~\ref{tab:real_world_inst_rb} and \ref{tab:real_world_inst_dup}, the performance on real instances is also much better than the worst-case guarantees, which justifies looking at random graphs for an indication of real-world performance. The results using both the random-bipartition method and the duplicating method are very similar in terms of the experimental performance-based ranking of the algorithms, with the graphs produced using the random-bipartition method being seemingly easier. The numbers achieved by \textsc{Ranking} and \textsc{SimpleGreedy} are pretty much identical when random-bipartition is used, while \textsc{Ranking} is significantly better when the duplicating method is used. \textsc{Ranking} still outperforms all non-greedy algorithms, but unlike in stand-alone graphs, there is always a greedy version of an algorithm that uses the type graph that performs just as well. In datasets \textit{Caltech36} and \textit{Reed98}, the ranking of Table~\ref{tab:algos_ranking} is maintained for non-greedy algorithms. In \textit{CE-GN} and \textit{CE-PG}, \textsc{ManshadiEtAl} and \textsc{BahmaniKapralov} outperform the rest. In \textit{beause} and \textit{mbeaflw} \textsc{ManshadiEtAl} is the best algorithm, while in \textit{beause} using the random-bipartition method, \textsc{BahmaniKapralov} comes a close second, outperforming \textsc{BrubachEtAl} and \textsc{JailletLu}.

The experimental performance-based ranking of the  non-greedy algorithms is generally consistent with their provable competitive ratios. \textsc{BrubachEtAl} and \textsc{JailletLu} perform very similarly and are usually on top, although on some graphs, \textsc{ManshadiEtAl} and \textsc{BahmaniKapralov} can outperform the rest. \textsc{FeldmanEtAl} almost always comes last (with a few exceptions, e.g. Molloy-Reed), but \textsc{FeldmanEtAl(g)} is often one of the best online algorithms. \textsc{Ranking} and \textsc{SimpleGreedy} get excellent results (with the exception of \textbf{UT}) and almost always outperform all non-greedy algorithms. After turning non-greedy algorithms into greedy algorithms, the ranking of the resulting algorithms is not very predictable. \textsc{FeldmanEtAl(g)} does quite well and its place in the ranking improves. \textsc{BrubachEtAl(g)} and \textsc{JailletLu(g)} can drop in rankings and become some of the worst greedy algorithms, while it is not unsual for \textsc{BahmaniKapralov(g)} to become the best. It is not straightforward to predict how an algorithm will behave after it turns greedy but the transformation appears to be very beneficial as the greedy versions significantly outperform non-greedy algorithms. \textsc{Category-Advice} seems to be a great algorithm to use in streaming models and when dealing with massive datasets. Even though it can improve substantially  over the \textsc{SimpleGreedy} solution, an additional pass (as done in \textsc{3-Pass}) does not seem to provide much benefit, as discussed before.

\section{Conclusion}
\label{sec:conclusion}
In this paper, we experimentally studied various online bipartite matching algorithms under the known i.i.d. input model with integral types. Type graphs that were used in our evaluations came from different sources, including random models of social networks, real-life networks, and stand-alone graphs that appeared previously in matching-related literature. Broadly speaking, algorithms under consideration can be split into two groups: simple algorithms that do not make use of the additional information (i.e., the type graph), and more complicated algorithms that often have a computationally intensive preprocessing step that tries to utilize the type graph for future predictions. The more complicated algorithms were developed and analyzed by researchers in the worst-case known i.i.d. setting so as to demonstrate more realistic performance bounds in contrast to the purely adversarial setting. These algorithms are often presented as being non-greedy to simplify the analysis. In contrast, most simple algorithms are naturally greedy. It is relatively easy to convert  complicated algorithms into greedy ones without hurting the  worst-case performance guarantee and without any significant  computational overhead. Thus, intuitively an algorithm for online bipartite matching can be viewed as consisting of two parts; namely, the complicated preprocessing part, and the greedy part. One of the main questions we try to answer in this work is how much each part is contributing to the competitive ratio on ``practical instances,'' where practical instances are modeled by the type graphs discussed above. It turns out that most of the work is done by the greedy part. In particular, the simple greedy algorithm tends to outperform all non-greedy versions, sometimes quite significantly. It also tends to perform comparable to the greedy versions of more complicated algorithms. In certain scenarios, the more complicated algorithms turned into greedy ones outperform the simple greedy algorithm, although it is questionable whether the performance boost is worth the extra computational effort in practice. In certain cases this overhead can become computationally intractable in practice (e.g., running \textsc{BrubachEtAl} on type graphs with millions of nodes).


There are many problems suggested by our work and many future directions are worth exploring. We list some of them here:


\begin{openproblem}
We conjecture that a practical study of online bipartite matching under the known i.i.d. with fractional types would result in very similar results and conclusions to what we observed with integral types. Does there exist a ``practical instance'' with fractional types that highlights the necessity to use more complicated algorithms?
\end{openproblem}

\begin{openproblem}
It is important to perform a similar evaluation on real-life data for online advertising, which is one of the main applications of online bipartite matching. Such data is proprietary and is not available to the public. Creating a public repository of such benchmarks would be a great contribution to the field on its own.
\end{openproblem}

\begin{openproblem}
In order to bridge the gap between theory and practice one needs to consider models other than worst-case. Known i.i.d. was the first step in this direction for online bipartite matching, since worst-case over type graphs allows for much better competitive ratios than worst-case over adversarial inputs. However, the area does not have to stop at known i.i.d. It is important to design and analyze new stochastic input models that better match practical inputs for certain application domains, e.g., online advertising.
\end{openproblem}

\begin{openproblem}
In an attempt at being fair and test all algorithms on same type graphs, we were limited to consider graphs with at most 1000 nodes due to prohibitive computational requirements of \textsc{BrubachEtAl}. One could perform a study on extremely large instances with millions or billions of nodes by excluding algorithms with too much preprocessing. We suspect that results of such a study would be similar to ours, and they would highlight the importance of using very simple greedy algorithms in large-scale applications. Would any of the complicated algorithms be able to handle such instances? Would the extra computation be worth it?
\end{openproblem}

\noindent{\bf Acknowledgements.} We thank Michael Kapralov for discussing \textsc{BahmaniKapralov} algorithm. Part of the work was done while the first author was at the Toyota Technological Institute at Chicago, and the last author was a postdoc at the University of Toronto.

\bibliographystyle{plain}
\bibliography{thesis}

\appendix

\end{document}